\documentclass[twocolumn]{aastex62}
\pdfoutput=1

\usepackage{amsmath}
\graphicspath{{./}{figures/}}
\received{May 29, 2020}
\revised{October 2, 2020}
\accepted{\today}
\submitjournal{ApJ}

\shorttitle{AGN Enhancement in Galaxy Interactions}
\shortauthors{Shah et al.}

\begin{document}

\title{Investigating the Effect of Galaxy Interactions on AGN Enhancement at $0.5<z<3.0$}

\correspondingauthor{Ekta A. Shah}
\email{eas7266@rit.edu}

\author[0000-0001-7811-9042]{Ekta A. Shah}

\affil{School of Physics and Astronomy, Rochester Institute of Technology, 84 Lomb Memorial Drive, Rochester NY 14623, USA}
\affil{LSSTC DSFP Fellow}

\author[0000-0001-9187-3605]{Jeyhan S. Kartaltepe}

\affil{School of Physics and Astronomy, Rochester Institute of Technology, 84 Lomb Memorial Drive, Rochester NY 14623, USA}

\author[0000-0001-6333-8090]{Christina T. Magagnoli}
\affil{School of Physics and Astronomy, Rochester Institute of Technology, 84 Lomb Memorial Drive, Rochester NY 14623, USA}

\author[0000-0002-1803-794X]{Isabella G. Cox}
\affil{School of Physics and Astronomy, Rochester Institute of Technology, 84 Lomb Memorial Drive, Rochester NY 14623, USA}

\author[0000-0002-9279-5351]{Caleb T. Wetherell}
\affil{School of Physics and Astronomy, Rochester Institute of Technology, 84 Lomb Memorial Drive, Rochester NY 14623, USA}

\author[0000-0002-8163-0172]{Brittany N. Vanderhoof}
\affil{School of Physics and Astronomy, Rochester Institute of Technology, 84 Lomb Memorial Drive, Rochester NY 14623, USA}

\author[0000-0003-2536-1614]{Antonello Calabro}
\affil{INAF OAR, via Frascati 33, Monte Porzio Catone 00078, Italy}

\author[0000-0003-3691-937X]{Nima Chartab}
\affil{Department of Physics and Astronomy, University of California, Riverside, 900 University Ave, Riverside, CA 92521, USA}

\author{Christopher J. Conselice}
\affil{Centre for Particle Theory and Astronomy, University of Nottingham, Nottingham NG7 2RD, UK}

\author[0000-0002-5009-512X]{Darren J. Croton}
\affil{Centre for Astrophysics \& Supercomputing, Swinburne University of Technology, P.O. Box 218, Hawthorn, Victoria 3122, Australia}

\author[0000-0002-6589-2017]{Jennifer Donley}
\affil{Los Alamos National Laboratory, P.O. Box 1663, Los Alamos, NM 87545, USA}

\author{Laura de Groot}
\affil{Department of Physics, The College of Wooster, 1189 Beall Avenue, Wooster, OH 44691, USA}

\author[0000-0002-6219-5558]{Alexander de la Vega}
\affil{Department of Physics and Astronomy, Johns Hopkins University, Baltimore, MD, 21218}

\author[0000-0001-6145-5090]{Nimish P. Hathi}
\affil{Space Telescope Science Institute, 3700 San Martin Dr., Baltimore, MD 21218, USA}

\author{Olivier Ilbert}
\affil{Aix Marseille Universit\'e, CNRS, LAM (Laboratoire d’Astrophysique de Marseille) UMR 7326, 13388, Marseille, France}

\author[0000-0003-4268-0393]{Hanae Inami}
\affil{Hiroshima Astrophysical Science Center, Hiroshima University,
1-3-1 Kagamiyama, Higashi-Hiroshima, Hiroshima 739-8526, Japan}

\author{Dale D. Kocevski}
\affil{Department of Physics and Astronomy, Colby College, Waterville, ME 04961, USA}

\author[0000-0002-6610-2048]{Anton M. Koekemoer}
\affil{Space Telescope Science Institute, 3700 San Martin Dr., Baltimore, MD 21218, USA}

\author[0000-0002-1428-7036]{Brian C. Lemaux}
\affil{Department of Physics \& Astronomy, University of California, Davis, One Shields Ave., Davis, CA 95616, USA}

\author{Kameswara Bharadwaj Mantha}
\affil{Department of Physics and Astronomy, University of Missouri-Kansas City, Kansas City, MO 64110, USA}

\author[0000-0001-5544-0749]{Stefano Marchesi}
\affil{INAF - Osservatorio di Astrofisica e Scienza dello Spazio di Bologna, Via Piero Gobetti, 93/3, 40129, Bologna, Italy}
\affil{Department of Physics and Astronomy, Clemson University,  Kinard Lab of Physics, Clemson, SC 29634, USA}

\author[0000-0001-5454-1492]{Marie Martig}
\affil{Astrophysics Research Institute, Liverpool John Moores University, 146 Brownlow Hill, Liverpool L3 5RF, UK}

\author{Daniel C. Masters}
\affil{IPAC, California Institute of Technology, 1200 E. California Blvd, Pasadena, CA 91125, USA}

\author{Elizabeth J. McGrath}
\affil{Department of Physics and Astronomy, Colby College, Waterville, ME 04961, USA}

\author{Daniel H. McIntosh}
\affil{Department of Physics and Astronomy, University of Missouri-Kansas City, Kansas City, MO 64110, USA}

\author[0000-0002-3430-3232]{Jorge Moreno}
\affil{Department of Physics and Astronomy, Pomona College, 333 N. College Way, Claremont, CA 91711, USA}

\author{Hooshang Nayyeri}
\affil{Center for Cosmology, Department of Physics and Astronomy, 4129 Reines Hall, University of California, Irvine, CA 92697, USA}

\author[0000-0002-4140-0428]{Belen Alcalde Pampliega}
\affil{Departamento de Física de la Tierra y Astrofísica, Facultad de CC Físicas, Universidad Complutense de Madrid E-2840 Madrid, Spain}

\author{Mara Salvato}
\affil{Max-Planck-Institut f\"{u}r extraterrestrische Physik (MPE),
Giessenbachstrasse 1, D-85748 Garching bei M\"{u}nchen, Germany}

\author{Gregory F. Snyder}
\affil{Space Telescope Science Institute, 3700 San Martin Dr., Baltimore, MD 21218, USA}

\author{Amber N. Straughn}
\affil{Astrophysics Science Division, NASA's Goddard Space Flight Center, Code 665, Greenbelt, MD 20771, USA}

\author[0000-0001-7568-6412]{Ezequiel Treister}
\affil{Instituto de Astrofisica, Facultad de Fisica, Pontificia Universidad Catolica de Chile, Casilla 306, Santiago 22, Chile}

\author{Madalyn E. Weston}
\affil{Department of Physics and Astronomy, University of Missouri-Kansas City, Kansas City, MO 64110, USA}

\begin{abstract}

Galaxy interactions and mergers are thought to play an important role in the evolution of galaxies. Studies in the nearby universe show a higher AGN fraction in interacting and merging galaxies than their isolated counterparts, indicating that such interactions are important contributors to black hole growth. To investigate the evolution of this role at higher redshifts, we have compiled the largest known sample of major spectroscopic galaxy pairs (2381 with $\Delta V <5000$\thinspace km s$^{-1}$) at $0.5<z<3.0$ from observations in the COSMOS and CANDELS surveys. We identify X-ray and IR AGN among this kinematic pair sample, a visually identified sample of mergers and interactions, and a mass-, redshift-, and environment-matched control sample for each in order to calculate AGN fractions and the level of AGN enhancement as a function of relative velocity, redshift, and X-ray luminosity. While we see a slight increase in AGN fraction with decreasing projected separation, overall, we find no significant enhancement relative to the control sample at any separation. In the closest projected separation bin ($<25 $\thinspace kpc, $\Delta V <1000$\thinspace km s$^{-1}$), we find enhancements of a factor of 0.94$^{+0.21}_{-0.16}$ and 1.00$^{+0.58}_{-0.31}$ for X-ray and IR-selected AGN, respectively. While we conclude that galaxy interactions do not significantly enhance AGN activity on average over $0.5<z<3.0$ at these separations, given the errors and the small sample size at the closest projected separations, our results would be consistent with the presence of low-level AGN enhancement.

\end{abstract}


\section{Introduction} \label{sec:intro}

Galaxy interactions and mergers play a crucial role in the evolution of galaxies. Studies based on observations in the nearby universe show that galaxy interactions have strong effects on the properties of galaxies, such as their morphology  \citep[e.g.,][]{lotz2008,darg2010, ellison2010}, star formation rates (SFRs) \citep[e.g.,][]{ellison2008, ellison2013a,patton2013}, and active galactic nuclei (AGN) activity \citep[e.g.,][]{alonso2007,woods2007,ellison2008,rogers2009,darg2010}.

Empirical relations such as the M$_{BH}-\sigma$ relation \citep{ferrarese2000,gebhardt2000,mcconnell2012} suggest that galaxies and their central supermassive black holes (SMBHs) evolve together. Hence, understanding the link between AGN/SMBHs and galaxy mergers is paramount to understanding the processes responsible for the co-evolution of galaxies and their SMBHs. There are two core questions related to the causal merger-AGN connection: (i) Do all galaxy mergers produce AGN? and (ii) Are mergers the primary trigger of AGN?

To answer the first question, studies compare the AGN activity of interacting and merging galaxies with isolated (non-interacting) galaxies. For low redshift ($0.01<z<0.20$) major galaxy pairs (stellar mass ratio $<4$) selected from the Sloan Digital Sky Survey (SDSS), \citet{ellison2013b} find a clear trend of increasing AGN excess (ratio of AGN fraction in paired galaxies compared to a control sample of isolated galaxies) with decreasing projected separation ($<40$\thinspace kpc). They measure the largest enhancement of $\sim$ 2.5 at the closest projected separation ($<$ 10 kpc). Numerous studies using a similar approach in the nearby universe find significant AGN enhancement in merging and/or interacting galaxies \citep{alonso2007, woods2007, ellison2011, satyapal2014, weston2017, ellison2019}.

For the second question, studies compare the merger and/or interaction fraction of an AGN sample with that of galaxies without AGN. More than 80\% of quasars (high luminosity AGN) in the nearby universe show signs of a recent or ongoing merger \citep{sanders1988a,sanders1988b,bannert2008,urrutia2008}. Similarly, \citet{koss2010} find a higher fraction of disturbed galaxies (18\% versus 1\%) and close pairs within 30\thinspace kpc (24\% versus 1\%) in \textit{Swift} BAT hard X-ray  moderate-luminosity AGN compared to normal galaxies at $z<0.5.$ However, \citet{ellison2019} show that about 63\% of optically-selected AGN host galaxies from SDSS do not show visual signs of disturbance and they do not have a companion galaxy within a 30\thinspace kpc projected separation, suggesting that recent interactions and mergers are not the primary trigger of optical AGN. They also show that almost 60\% of mid-IR selected AGN show signs of disturbances; hence, interactions play a significant role in feeding AGN, and obscured AGN are more likely to be triggered via mergers.

At high redshift, the merger-AGN connection is even more controversial. Using a sample of 562 spectroscopic galaxy pairs (mass ratio $<10$ and $0.25<z<1.05$), \citet{silverman2011} find a higher ($\times1.9$) AGN fraction in paired galaxies at projected separation less than 75\thinspace kpc compared to a control galaxy sample. \citet{lackner2014} apply an automated method of identifying mergers by median-filtering of the high-resolution COSMOS \textit{Hubble Space Telescope} (HST) images \citep{koekemoer2007} to distinguish two concentrated galaxy nuclei at small separations. They use this method to identify late-stage mergers at $0.25<z<1.0$ and use X-ray observations to identify AGN. They find higher ($\times2$) X-ray selected AGN activity in their late-stage merger sample compared to a mass- and redshift matched control sample. \citet{treister2012} find a luminosity-dependence of the merger-AGN connection at all redshifts ($0<z<3$), showing that the merger fraction in AGN increases from less than $\sim$25\% for low-moderate luminosity AGN  ($\sim10\%$ for all AGN) to $\sim$70-80\% for the highest luminosity AGN ($L_{\rm{bol}}>10^{46}$\thinspace erg s$^{-1}$). This higher merger fraction in high luminosity AGN is absent in other studies  \citep[e.g.,][]{villforth2016, hawlett2017}. Most studies based on low or intermediate luminosity AGN  \citep[e.g.,][]{grogin2005,schawinski2011,kocevski2012} do not find a higher merger fraction in AGN at high redshift compared to non-AGN.

\citet{schawinski2012} study heavily obscured quasars at $z\sim2$ and find a very low merger fraction in these AGN hosts, concluding that most of them are disks and not mergers. However, \citet{donley2018} show that about 75\% of luminous, heavily obscured IR-only AGN (not X-ray detected) in CANDELS/COSMOS are potentially late-stage major mergers. \citet{kocevski2015} find that $\sim$22\% of heavily obscured AGN at $z\sim1$ show signs of interaction or merger compared to  unobscured AGN ($\sim$8\%). Hence, different types of AGN might be triggered by different processes.

Part of this discrepancy could be due to the different methods used to identify galaxy merger and/or galaxy pair samples, corresponding control samples, and the identification of AGN in galaxies. Most of the studies use one of two methods to identify mergers/interactions: (i) using morphological signs of disturbances such as tidal tails, double nuclei, and tidal bridges, and (ii) identifying close pairs based on either spectroscopic or photometric redshifts.  The first method is challenging at high redshift, as observable merger signatures are difficult to identify because of their low surface brightness. The second method, however, can provide a larger and more complete sample of interacting galaxies. Furthermore, it also includes fly-bys that may not eventually merge, but could still have an impact on fueling AGN activity. To identify and confirm interacting galaxy pairs, high spectroscopic completeness is required. One can use photometric redshifts to select pairs, but the relatively large uncertainties on photometric redshifts increase the likelihood of a given pair being a chance projection along the line of sight rather than being physically associated.

The discrepancy could also be due to the use of different methods to identify AGN, such as the detection of broad emission lines, using X-ray (or radio) luminosity thresholds to identify X-ray (radio) AGN, emission line flux ratios to distinguish AGN-dominated galaxies from star formation dominated galaxies, and IR broadband colors to identify galaxies with a strong power law slope in the mid-infrared. Each of these methods traces different physical components of AGN (such as the accretion disk, dusty torus, radio lobes, jets, emission line regions, etc.). The identification of AGN is therefore not consistent among all these methods.  It is possible that an AGN might be identifiable at different wavelengths at different stages of the merger process or the AGN duty cycle, which could lead to different selection techniques resulting in different AGN fractions.

Although most massive galaxies have a SMBH at their center, only a relatively small fraction of SMBHs are actively accreting. Simulations of gas-rich galaxy mergers in the local universe show these events can provide the torques necessary for reducing most of the angular momentum ($\sim$99\%) of gas in the galaxy, funneling gas inflows towards the nuclear region ($\sim$1\thinspace kpc), ultimately triggering AGN activity at 1\thinspace pc scales \citep[e.g.,][]{barnes1991,mihos1996,dimatteo2008,hopkins2009}. However, the properties of these gas inflows (mass, size, shape, strength, etc.) and their propagation could be sensitive to the gas fraction and gas distribution in the galaxies.

The average gas fraction of galaxies changes significantly with redshift. At $z\sim2$, the gas fraction in massive spiral galaxies can be $\sim$50\%, compared to $\sim$10\% at $z\sim0$ \citep{daddi2010,tacconi2010,scoville2014}. Furthermore, the distribution of gas is very clumpy, and its average velocity dispersion is higher ($\sigma\sim40$\thinspace km s$^{-1}$) in high redshift galaxies compared to that ($\sigma \thinspace \sim \thinspace 10 $\thinspace km s$^{-1}$) in low redshift galaxies \citep{stott2016}. While the abundance of gas in high redshift galaxies might make it easier to form gas inflows through interactions, the high turbulence and velocity dispersion throughout the galaxy might weaken the propagation of inflows. Results of some simulations show significantly weaker gas inflows in high-redshift galaxy mergers compared to low-redshift galaxy mergers \citep[e.g.,][]{fensch2017,dimatteo2008}. Hence, the efficiency of galaxy interactions in enhancing AGN activity may change substantially with redshift \citep{mcalpine2020}.

Observing the evidence for this effect requires deep multiwavelength observations of a large sample of galaxy pairs and mergers over a wide redshift range. Using the multiwavelength observations and dedicated spectroscopic surveys in the CANDELS \citep{grogin2011,koekemoer2011} and COSMOS \citep{scoville2007} fields, we generated the largest known sample of 2381 spectroscopic galaxy pairs with a relative line of sight velocity less than 5000\thinspace km s$^{-1}$ undergoing major (stellar mass ratio of primary to secondary $<4$) galaxy interactions at $0.5<z<3.0$. We also compiled a sample of mass-, redshift-, and environment-matched isolated control galaxies. We use X-ray and IR observations to identify AGN and compare the AGN fraction in paired and control galaxies to estimate interaction-induced AGN enhancement in paired galaxies.

The layout of this paper is as follows. We describe the survey data used and our spectroscopic observations in Section \ref{sec:data}. In Section \ref{sec:sampleselection}, we discuss the methods used to generate our galaxy pair, visually identified mergers, and control samples. In Section \ref{sec:agnidentification}, we identify X-ray and IR-selected AGN. We estimate the AGN fraction and present our results on AGN enhancement for the spectroscopic galaxy pair and visually identified samples in Section \ref{sec:agn-enh-def} and Section \ref{sec:agn-enh-vis}, respectively. We discuss our results in Section \ref{sec:discussion} and summarize in Section \ref{sec:summary}. Throughout this work, we use a standard $\Lambda$CDM cosmology with $H_0=70$\thinspace km s$^{-1}$ Mpc$^{-1}$, $\Omega_\Lambda=0.7$, and $\Omega_M= 0.3$. All magnitudes are given in the observed AB system and mass values of the galaxies correspond to their stellar masses unless stated otherwise.

\section{Data} \label{sec:data}
	We use deep multiwavelength CANDELS \citep[PIs: S. Faber and H. Ferguson;][]{grogin2011,koekemoer2011} and COSMOS \citep{scoville2007} observations for this study. Due to the extensive multiwavelength photometric and spectroscopic observations available in these fields, they provide a statistically robust and complete sample of massive galaxies out to redshift $\sim3$, required for our study.

	CANDELS is a Multi-Cycle \textit{HST} Treasury program spanning an area of $\sim960$ arcmin$^2$. It consists of two types of surveys covering five different fields on the sky: (i) the CANDELS/Deep Survey, covering  $\sim125$ arcmin$^2$ within the Great Observatories Origins Deep Survey \citep[GOODS;][]{giavalisco2004} North (GOODS-N) and South (GOODS-S) fields, and (ii) the CANDELS/Wide Survey covering portions of GOODS-N, GOODS-S, the Extended Groth Strip \citep[EGS;][]{davis2007}, the Cosmic Evolution Survey \citep[COSMOS;][]{scoville2007}, and the  UKIDSS Ultra-Deep Survey \citep[UDS;][]{lawrence2007}. All of these five fields were observed using near-IR filters F160W and F125W on \textit{HST}/Wide Field Camera 3 (WFC3) and F606W and F814W on \textit{HST}/Advanced Camera for Surveys (ACS).

	 COSMOS is the largest ($\sim2$ deg$^2$) contiguous area HST survey \citep{koekemoer2007,scoville2007}, with coverage in ACS/F814W and a wealth of multiwavelength observations across the spectrum. The large area of COSMOS enables statistical studies of large samples, and in particular, allows for detailed analysis of the surrounding environment of galaxies and its impact on their evolution. In addition to the CANDELS observations of a small portion of COSMOS mentioned above, we also make use of observations across the full field in our analysis.

\subsection{Photometry and Derived Physical Quantities}

The source catalogs in the CANDELS fields were generated using the source detection algorithm \textsc{Source Extractor} \citep{bertin1996} applied to the F160W (H-band) 1 1/3 orbit depth CANDELS mosaic image for each field. We use the observed-frame multiwavelength (UV to Near-IR) photometric catalogs  produced by \cite{nayyeri2017}, \cite{guo2013}, \citet{barro2019}, \cite{galametz2013}, and \cite{stefanon2017} for the COSMOS, GOODS-S, GOODS-N, UDS, and EGS fields, respectively. The  final catalogs were compiled by combining multiwavelength observations with different spatial resolutions using the template-fitting method TFIT \citep{laidler2007,lee2012}, which provides uniform photometry across different filters. These catalogs also contain the photometric redshift values of the galaxies which were estimated using the method described by \cite{dahlen2013}. This method combines the posterior probability distribution of photometric redshifts from several different codes and template sets used for spectral energy distribution (SED) fitting and chooses the median of the peak redshifts of the different Probability Distribution Functions (PDFs) as the best available photometric redshift.

To estimate the stellar masses of the galaxies, ten different groups within the CANDELS team  fit the observed multiwavelength photometric observations with a set of SED templates with different stellar populations for a given redshift \citep{santini2015,mobasher2015}. These masses were then combined by computing the average of the posterior PDF and choosing the median of the estimates as the stellar mass for a given object. Each group used their preferred fitting code, assumptions, priors, and parameter grid to determine the stellar mass using the same photometry.

For galaxies in the full $\sim2$ deg$^2$ COSMOS field, we used the photometric catalog compiled by \cite{laigle2016}. The catalog contains photometry in 30 bands for more than half a million galaxies spanning a large redshift range up to $z\sim6$ and their precise photometric redshifts and stellar masses.  The source detection for COSMOS was also carried out using \textsc{Source Extractor} \citep{bertin1996}. The final detection image was generated by combining NIR images from UltraVISTA with the optical broad band observations from Subaru. To estimate photometric redshifts Laigle et al.\ used the NUV band observations from \textit{GALEX}, u$^*$ band data from the Canada-France-Hawaii Telescope (\textit{CFHT}/MegaCam), as well as 6 broad bands ({\it B, V, g, r, i, z$^++$}), 12 medium bands (\textit{IA427, IA464, IA484, IA505, IA527, IA574, IA624, IA679, IA709, IA738, IA767, and IA827)}, and two narrow bands (NB711, NB816) obtained using \textit{Subaru} SuprimeCam. SED fits were performed using the code {\sc LePhare}\footnote{\url{http://www.cfht.hawaii.edu/~arnouts/LEPHARE/lephare.html}} \citep{arnouts2002,ilbert2006}, which uses a wide range of templates of star-forming and quiescent galaxies from \citet{bruzual2003}. Extinction was added as a free parameter using the following extinction laws: \cite{calzetti2000}, \citet{prevot1984}, and \citep{fitzpatrick1986}. The contribution of emission lines was also considered using an empirical relation between the UV radiation and the emission line flux values \citep{ilbert2009}.

\cite{laigle2016} use {\sc LePhare} to estimate the stellar masses of the observed galaxies using a  \cite{chabrier2003} IMF, two metallicities (solar and half-solar), emission lines \citep{ilbert2009}, two attenuation curves \citep{calzetti2000,arnouts2013}, an exponentially declining and delayed star formation history, and a library of synthetic spectra generated based on the Stellar Population Synthesis model of \cite{bruzual2003}. For the area where the COSMOS and CANDELS-COSMOS survey fields overlap, we use the CANDELS catalogs rather than COSMOS because the WFC3-selected catalog has higher angular resolution and allows us to select pairs at closer separations.

As the above-mentioned stellar masses were mostly estimated using photometric redshifts, we recompute the stellar masses of our galaxy pairs and control galaxies using the spectroscopic redshifts with the SED fitting tool Multiwavelength Analysis of Galaxy Physical Properties \citep[{\sc MAGPHYS};][]{dacunha2008} using the photometry described above and the \cite{bruzual2003} stellar population libraries. We choose this code as it efficiently measures stellar masses and star formation rates for high redshift galaxies in a self-consistent manner. We compare our new masses with the original ones and find that they are consistent for cases where the redshift did not change. In this paper, we use the stellar masses we recomputed with the spectroscopic redshifts unless stated otherwise. In particular, we use these stellar masses to define the final spectroscopic galaxy pair and control samples described in detail in Section \ref{sec:sampleselection}. The star formation rates will be discussed in a subsequent paper.

\subsection{X-ray Observations}\label{subsec:xraydata}
In order to identify AGN based on X-ray emission, we used deep \textit{Chandra} X-ray observations in UDS \citep{kocevski2018}, GOODS-S \citep{xue2011}, GOODS-N  \citep{alexander2003}, EGS \citep{laird2009,nandra2015}, and COSMOS \citep[Chandra COSMOS-Legacy Survey;][]{elvis2009,civano2016,marchesi2016} with the full band ($0.5-10$\thinspace keV) limiting fluxes of $4.4\times10^{-16}$, $3.2\times10^{-17}$, $2.5\times10^{-17}$, $1.5\times10^{-16}$, and $9\times10^{-16}$ erg s$^{-1}$ cm$^{-2}$, respectively.

\subsection{\textit{Spitzer Space Telescope} Infrared Observations}\label{subsec:irdata}
To identify infrared-selected AGN (IR AGN), we used observations obtained with the four Infrared Array Camera (IRAC) channels ($3.6\thinspace\mu$m, $4.5\thinspace\mu$m, $5.8\thinspace\mu$m, $8.0\thinspace\mu$m) on the \textit{Spitzer Space Telescope} in all the fields: COSMOS \citep{sanders2007,ashby2013,laigle2016}, UDS \citep{,ashby2013,ashby2015}, GOODS (N-S) \citep{dickinson2003,giavalisco2004,ashby2013}, and EGS \citep{barmby2008,ashby2015}.

\subsection{Spectroscopic Observations}

 In this study, we used all known existing spectroscopic redshifts in the CANDELS and COSMOS fields, as compiled by each of the teams and assigned quality flags on a consistent system. We combined these redshifts with our own measured redshifts from our observations obtained using the  DEep Imaging Multi-Object Spectrograph (DEIMOS) on the \textit{Keck II} telescope, described below.

For the GOODS-S field, we use spectroscopic redshift measurements obtained using observations from the \textit{Very Large Telescope (VLT)}/ Visible Multi-Object Spectrograph (VIMOS) \citep{lefevre2004,ravikumar2007,balestra2010, lefevre2013,mclure2018}, \textit{VLT}/FORS1 (FORS: the visual and near UV FOcal Reducer and low dispersion Spectrograph) and \textit{VLT}/FORS2 \citep{daddi2004, szokoly2004,vanderwel2004,mignoli2005,vanzella2008,popesso2009,vanzella2008,vanzella2009,balestra2010,kurk2013,pentericci2018}, \textit{VLT}/the Multi Unit Spectroscopic Explorer (MUSE) \citep{inami2017,urrutia2019}, \textit{HST}/WFC3-IR grism spectroscopy  \citep{ferreras2009,morris2015, momcheva2016}, \textit{Gemini}/Gemini Multi-Object Spectrographs (GMOS) \citep{roche2006}, \textit{Keck I}/Multi-Object Spectrometer For Infra-Red Exploration (MOSFIRE) \citep{kriek2015}, \textit{Keck II}/DEIMOS \citep{silverman2010,cooper2012c}, and the \textit{Anglo-Australian Telescope (AAT)}/LDSS++ spectrograph \citep{croom2001}.

For the GOODS-N field, we use spectroscopic redshift values estimated using observations from \textit{HST}/WFC3-IR grism spectroscopy  \citep{ferreras2009,momcheva2016}, \textit{Keck I}/MOSFIRE and Low Resolution Imaging Spectrometer (LRIS) \citep{cowie2004,reddy2006, barger2008,kriek2015,wirth2015}, \textit{Keck II}/DEIMOS  \citep{wirth2004, cowie2004,barger2008,cooper2011}, and \textit{Subaru Telescope}/Multi-Object Infrared Camera and Spectrograph (MOIRCS) \citep{yoshikawa2010}.

The spectroscopic redshift values we use for the EGS field are based on spectroscopic observations acquired using \textit{Keck I}/MOSFIRE and LRIS \citep{coil2004,masters2019,kriek2015}, \textit{Keck II}/DEIMOS \citep{masters2019,cooper2012b,newman2013}, and \textit{HST}/WFC3-IR grism spectroscopy \citep{momcheva2016}.

For the UDS field, we use spectroscopic redshift estimates based on observations from \textit{HST}/WFC3-IR grism spectroscopy \citep{morris2015,momcheva2016}, \textit{VLT}/VIMOS and FORS2 \citep{bradshaw2013,pentericci2018}, \textit{Keck I}/MOSFIRE and LRIS \citep{kriek2015,masters2019}, \textit{Keck II}/DEIMOS \citep{masters2019}, and \textit{VLT}/VIMOS \citep{mclure2018,scodeggio2018}.

For the COSMOS field, we use spectroscopic redshifts estimated from observations obtained using \textit{VLT}/VIMOS \citep{lilly2007,tasca2015,lefevre2015,vanderWel2016,straatman2018}, \textit{VLT}/FORS2 \citep{comparat2015,pentericci2018}, \textit{Keck}-I/MOSFIRE and LRIS \citep{kriek2015,masters2019}, \textit{Keck II}/DEIMOS \citep{capak2004,kartaltepe2010, hasinger2018,masters2019}, \textit{MMT}/Hectospec spectrograph \citep{damjanov2018}, \textit{Subaru}/MOIRCS \citep{onodera2012}, \textit{Subaru}/FMOS (Fiber multi-Object Spectrograph) \citep{silverman2015,kartaltepe2015}, \textit{HST}/WFC3-IR grism spectroscopy \citep{krogager2014,momcheva2016}, and \textit{Magellan (Baade) telescope}/Inamori Magellan Areal Camera and Spectrograph (IMACS) \citep{trump2009, coil2011}.

We also use spectroscopic observations obtained using \textit{Gemini}/GMOS (I. Cox et al., in preparation) and \textit{Keck I}/MOSFIRE (B. Vanderhoof et al., in preparation) for the UDS, COSMOS, and GOODS-S fields.

\subsection{\textit{Keck II} DEIMOS Observations} \label{sec:deimosobs}

Apart from the above mentioned spectroscopic observations, we also include spectra of galaxies observed with DEIMOS (PI: J. Kartaltepe). DEIMOS is an optical (4000\thinspace\AA\ -- 10500\thinspace\AA) multiobject imaging spectrograph mounted on the \textit{Keck II Telescope} \citep{faber2003}. In a single exposure, DEIMOS can simultaneously take spectroscopic observations of more than 100 galaxies, covering a wide spectral range of up to 5000\thinspace\AA\  with a high spectral resolution ($R\sim2000$ with the 600 l/mm grating). The user can specify the length, width, position, and position angle (PA) of individual slits. These characteristics make DEIMOS one of the best instruments for obtaining spectroscopic observations, and hence estimating spectroscopic redshifts, of a large number of galaxies over a wide area.

For the DEIMOS observations, we select galaxy pair candidates using stellar masses and photometric redshifts from the CANDELS team-derived catalogs using the pair selection criteria described in Section \ref{sec:pairselection}. From these galaxy pair candidates, we select those without spectroscopic redshift values available at the time, to generate a target candidate list. In this list, we also include other (e.g., \textit{Herschel Space Observatory} detected) galaxies without spectroscopic redshifts as fillers. We assign a higher priority to the galaxy pair candidates (primary targets) and lower to the filler galaxies (secondary targets).

To design DEIMOS slitmasks, we use the \textsc{dsimulator}\footnote{\url{https://www2.keck.hawaii.edu/inst/deimos/dsim.html}} slitmask software, which creates the final target list from the target candidate list. We choose positions and PAs of the masks and corresponding slits to cover both members of the galaxy pairs at the smallest separations if possible, or to follow the major axis of the galaxy. We created a total of twelve masks for the observations in the CANDELS-COSMOS field and nine masks for the CANDELS-UDS field with $\sim100$ targets per mask.

We observed the CANDELS-COSMOS and CANDELS-UDS fields over two observing runs -- December 16 $\&$ 17, 2014, and January 30, 2017. There were clouds throughout the 2014 run, which affected the data quality, so only the brightest galaxies were detected. However, the weather was clear with seeing $\sim$ 0.5" throughout the observation night in 2017. For wavelength calibration, we carried out observations of the Ne, Ar, Kr, and Xe arc lamps. During the observation run in 2014, we observed eight slitmasks for each of the two fields, and during the observation run in 2017, we observed four slitmasks for the CANDELS-COSMOS field and one slitmask for the CANDELS-UDS field. We used the 600ZD grating on the DEIMOS instrument for these observations. Each mask was observed for $\sim100$ mins.

We reduced the data using the publicly available spec2d IDL pipeline created for the DEIMOS instrument \citep{newman2013,cooper2012a}.  The spec2d pipeline extracts sources and their corresponding sky-subtracted and calibrated one-dimensional (1D) and two-dimensional (2D) spectra. In some cases, we obtained more than one spectrum (targeted source and serendipitous source) for a given slit. For some of them, the serendipitous source was the companion galaxy of the corresponding pair candidate. For other cases, the serendipitous source(s) was (were) just a background/foreground source(s).

\begin{figure}[ht]
    \centering
    \hspace*{-0.2in}
    \includegraphics[scale=0.37,trim = 1cm 1cm 1cm 1.3cm,clip=true]{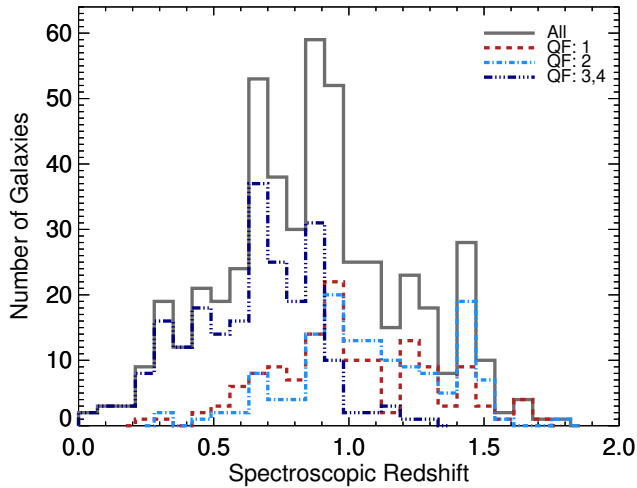}
    \caption{Distribution of spectroscopic redshift values obtained from DEIMOS observations in UDS and COSMOS (gray line) with low quality flag of 1 (dashed red line), 2 (dot-dashed light blue line), and high quality flag 3 or 4 (dot-dot-dot-dashed navy line). Most of the $z<1$ redshifts are of high quality since multiple bright lines are often observed while at $z>1$, only one strong line is typically seen, and therefore assigned a quality flag of 1 or 2.
    Note the spike at $z\sim 0.9$, which corresponds to several overdensities in both fields between $z\sim 0.9$ and 1.
    }
    \label{fig:deimz}
\end{figure}

For the measurement of spectroscopic redshifts, we used the SpecPro software package \citep{masters2011} with built-in spectroscopic templates for galaxy emission and absorption features.  We visually overlaid spectroscopic templates on the common emission and absorption features of the 1-D and 2-D observed spectra and used photometric redshifts as initial guess values. We estimated the spectroscopic redshift by shifting the emission templates along the wavelength axis until their emission and absorption features best match with the observed features. We defined four flags corresponding to the quality of the spectroscopic redshift value, consistent with the quality flags used by the CANDELS and COSMOS team spectroscopic compilations. Quality flag 1, 2, 3, and 4 corresponding, respectively, to one spectral line with low signal to noise ratio (SNR), one spectral line with high SNR, multiple spectral lines with low SNR, several spectral lines with high SNR. This scheme follows a simplified version of the flags defined by the zCOSMOS survey  \citep{lilly2009}. In the case where only one emission line was detected, we assume that it corresponded to the brightest line nearest the photometric redshift.

For the CANDELS-UDS field, we estimated spectroscopic redshifts for a total of 243 galaxies, out of which 105 have a high quality flag of 3 or 4, and 138 have a low quality flag of 1 or 2. For the CANDELS-COSMOS field we estimated spectroscopic redshifts for a total of 261 galaxies with 118 redshift values with a high quality flag (3,4) and 143 redshift values with a low quality flag (1,2). We present the spectroscopic redshift distribution (gray line) of galaxies observed with DEIMOS in Figure~ \ref{fig:deimz} subdivided into low quality flag 1 (dashed red line), quality flag 2 (dot-dashed light blue line) and
high quality flag 3 and 4 (dot-dot-dot-dashed navy line) quality flags. The distribution shows that most of the low redshift ($z<1$) and high redshift ($z>1$) estimates are dominated by high quality flags and low quality flags, respectively. This is mainly due to multiple bright lines observed for most of low redshift galaxies and only one bright line observed for most high redshift galaxies.

To summarize, we use the source positions, photometric redshifts, and stellar masses from the CANDELS and COSMOS photometric catalogs to identify galaxy pair candidates for targeting with our DEIMOS observations. We use the new spectroscopic redshifts, along with the existing spectroscopic redshifts gathered from the literature to recompute the stellar masses as described above, and use those new stellar masses throughout our analysis.

\section{Sample Selection} \label{sec:sampleselection}
  This section describes the criteria we use to generate (i) the spectroscopic galaxy pair sample, (ii) the visually identified-interacting galaxy and merger sample, and (iii)  the corresponding mass-, redshift-, and environment-matched isolated (control) galaxy sample using the CANDELS and COSMOS survey observations. Since AGN activity strongly depends on the stellar mass, redshift, and environment of a galaxy, in order to isolate the effect of interactions and mergers, we control for these variables by generating a mass-, redshift-, and environment-matched control sample corresponding to the galaxy pair sample.

\subsection{Pair Selection} \label{sec:pairselection}
We combine the photometric and spectroscopic catalogs in the COSMOS and CANDELS fields described above to obtain the coordinates, stellar masses, and the best spectroscopic redshifts for galaxies in each field. We only use spectroscopic redshifts with quality flag greater than one based on the above mentioned scheme for both the literature compilations and our DEIMOS observations.  We only consider massive galaxy pairs undergoing  major galaxy interactions by restricting the stellar mass of both galaxies in a pair to be greater than 10$^{10}$\thinspace M$_{\odot}$ and the stellar mass ratio of primary to secondary galaxy (less massive of the two galaxies in a pair) to be less than four, consistent with the typical values used in the literature \citep[e.g.,][]{ellison2013b,mantha2018}. Since the mass completeness at high redshift differs among the different CANDELS and COSMOS fields, in order to be consistent we constrain the redshift of paired galaxies to be less than three since all of the fields are complete down to 10$^{10}$ M$_{\odot}$ at this redshift. As the focus of this study is on high redshift interactions, and for $z<0.5$ each of the CANDELS fields contains a small volume, we restrict the spectroscopic redshift of the paired galaxies to be greater than 0.5. Ideally, we would measure the three-dimensional separation between galaxies to select the companion for a galaxy. However, in reality, we can only estimate the projected separation of galaxies. We calculate the projected physical separation of the two galaxies in a pair by using their angular separation and average spectroscopic redshift. To constrain the line of sight separation, we use the relative radial velocities obtained using the spectroscopic redshifts of the galaxies.

We use the following criteria to generate the sample of massive spectroscopic galaxy pairs undergoing major galaxy interactions:

\begin{enumerate}

\item \textit{Redshift limit}: The spectroscopic redshift of both of the galaxies in a pair has to be between 0.5 and 3.0.

\item \textit{Mass limit}: The stellar mass of both of the galaxies has to be greater than 10$^{10}$\thinspace M$_{\odot}$.

\item \textit{Stellar mass ratio}:  The stellar mass ratio of the primary to the secondary galaxy has to be less than four.

\item \textit{Relative line of sight velocity}: Companions are required to have their relative line of sight velocity (obtained using their spectroscopic redshifts) within 5000\thinspace km s$^{-1}$. This is an intentionally large relative velocity cut that enables us to test for the effect of different cuts. We explore the effect of using a $\Delta V < 500$, 1000, and 5000\thinspace km s$^{-1}$ selection throughout our analysis.

\item \textit{Projected separation}: We require the projected separation between companions to be less than 150\thinspace kpc.

\end{enumerate}

\begin{figure}[ht]
    \centering
    \hspace*{-0.2in}
    \includegraphics[scale=0.37,trim = 1cm 1cm 1cm 1.3cm,clip=true]{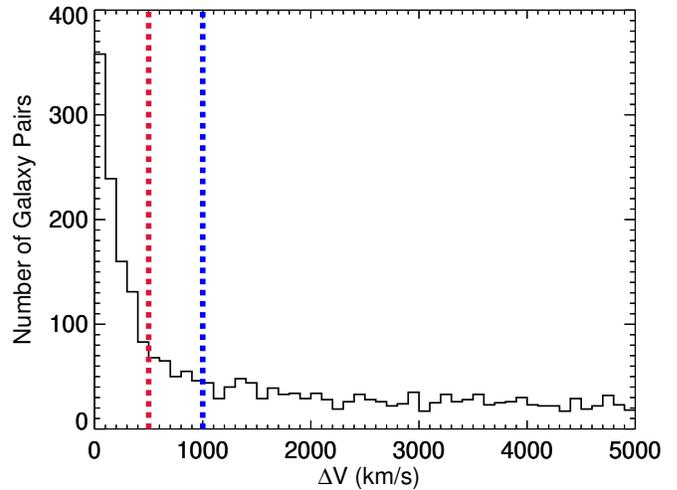}
    \caption{Line of sight relative velocity distribution of our sample of 2381 galaxy pairs with $\Delta V<5000$\thinspace km s$^{-1}$, with vertical lines highlighting the cuts of $\Delta V<1000$\thinspace km s$^{-1}$ (blue) and $\Delta V<500$\thinspace km s$^{-1}$ (red) used throughout this paper. The sharp peak at very small velocities indicates that the majority of these pairs are likely to be interacting.}
    \label{fig:vel_distr}
\end{figure}

\begin{figure}[ht]
    \centering
    \hspace*{-0.2in}
    \includegraphics[scale=0.37,trim = 1cm 1cm 1cm 1.3cm,clip=true]{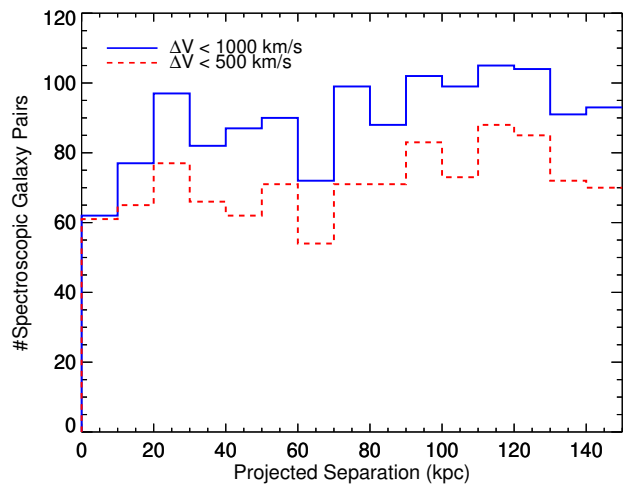}
    \caption{Projected separation distribution of galaxy pairs with $\Delta V<1000$\thinspace km s$^{-1}$ (blue) and $\Delta V<500$\thinspace km s$^{-1}$ (red). Note that while the overall distribution of the sample is relatively uniform, there is a dearth of pairs at the closest separations ($<10$\thinspace kpc), where close pairs are hardest to resolve.}
    \label{fig:sep_distr}
\end{figure}

To explore the effects of interactions as a function of the projected separation of a galaxy pair, we intentionally include potentially merging systems as well as pairs that are interacting/have interacted in the past but are not going to necessarily merge (they could still have been affected by the interaction). Hence, we want to cover a wide range of separation and relative velocity difference. While most studies consider the maximum projected separation of a galaxy pair to be $\sim80-100$\thinspace kpc \citep[e.g.,][]{patton2011,scudder2012,ellison2013a}, there are some studies that show that galaxy interactions can have effects on galaxy pairs with projected separation of up to 150\thinspace kpc \citep[e.g.,][]{patton2013}. Therefore, we restrict the maximum projected separation of galaxy pairs to 150\thinspace kpc.

 We present a total sample of 2381 spectroscopic major galaxy pairs satisfying all the conditions mentioned above. The relative velocity distribution of galaxy pairs satisfying all the criteria is shown in Figure~\ref{fig:vel_distr}. To maximize the chances of galaxies being physically associated and therefore the possibility of interaction, and to explore the effects of using different velocity cuts, we also apply more restrictive cuts to the relative velocity difference of less than 500\thinspace km s$^{-1}$ (1066 pairs) and 1000\thinspace km s$^{-1}$ (1345 pairs) and explore the effect of using different velocity cuts in our results. Table~\ref{table:int-gal-numbers} shows the number of galaxy pairs in each field satisfying all criteria.

 The projected separation distributions of these galaxy pair samples are shown in Figure~\ref{fig:sep_distr}, which is fairly uniform at separations greater than 20\thinspace kpc. There are relatively few systems in the smallest projected separation bin ($<10$\thinspace kpc). The minimum separation among the pairs in our sample is 4.4\thinspace kpc. Even with HST resolution, systems at closer separations are difficult to resolve at high redshift. Given the redshift range of our sample, the physical separation that we can resolve does not vary much with redshift. At closer separations, some pairs might be blended in our photometric measurements but still able to be detected visually. Such systems are described in the next section.

\begin{deluxetable}{cccc}

\tablewidth{0pt}
\tablecaption{Number of Major Spectroscopic Galaxy Pairs in Each Field}
\tablehead{\colhead{Field} & \colhead{}  & \colhead{\# Galaxy Pairs}  & \colhead{}\\
\colhead{}  & \colhead{$\Delta V<5000$}& \colhead{$\Delta V<1000$}  & \colhead{$\Delta V<500$}}
\startdata
COSMOS & 1802 & 1008 & 806\\
UDS & 127 & 72 & 52\\
GOODS-N & 82 & 44 & 37\\
GOODS-S & 211 & 140 & 110\\
EGS & 159 & 81 & 61\\
\hline
Total & 2381 & 1345 & 1066\
\enddata
\tablecomments{$\Delta V$ denotes relative line-of-sight velocity in km\thinspace s$^{-1}$.}

\label{table:int-gal-numbers}
\end{deluxetable}

\subsection{Visually Identified Interactions and Mergers}\label{sec:visual_int_mer}

To investigate different stages of the galaxy merger process, we also selected a subsample of visually identified interacting galaxies and mergers using the classification scheme and catalog of \citet{kartaltepemor15}. As mentioned above, the number of spectroscopic galaxy pairs with projected separation less than 10\thinspace kpc is limited in our sample as it is hard to resolve two galaxies with small separation in a pair at high redshift. However, pairs at these separations are more likely to show morphological signatures of interaction and less likely to be chance projections. Therefore, we include visually identified pairs as well as mergers that have coalesced into a single system in order to span the full range of physical separations and merger stages. A caveat to using the visually identified sample is that the observability of the morphological signs of mergers and interactions can strongly depend on different properties of the merging galaxies such as their morphological types, stellar masses and stellar mass ratio, redshift, gas fraction, orbital parameters of the merger, as well as observational factors such as the image depth, observed wavelength, viewing angle, etc. Hence, this sample does not represent a complete sample of interactions and mergers.

\begin{figure}[b]
\hspace*{-0.2in}
    \includegraphics[trim = 1.3cm 3cm 1.3cm 3cm,clip=true, scale=0.72]{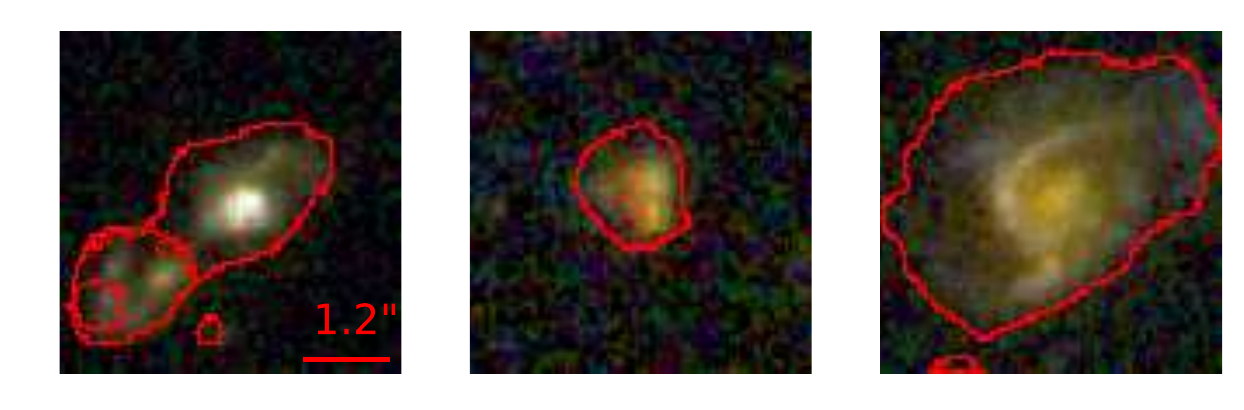}
    \caption{\textit{HST} F606W, F125W, and F160W composite images of an example of a visually identified non-blended interaction (left), a blended interaction (center), and a merger (right). The red contours show the outlines of the segmentation map. All the images are the same angular size and have a $1.2"$ scale bar. Note that each of these galaxies has observable tidal tails and disturbed morphology.}
    \label{fig:late_stage_mer}
\end{figure}

 \citet{kartaltepemor15} produced a visual classification catalog for all galaxies with $\textit{H}<24.5$ in the CANDELS fields, covering $\sim50,000$ galaxies in total. Each galaxy was visually classified by at least three individual classifiers.  In order to construct a sample of high confidence galaxy interactions and mergers, we selected galaxies where $\geq 2/3$ of all classifiers agreed that the galaxy was involved in an interaction or a merger, with additional cuts as described below.  A full catalog of galaxy mergers and interactions, along with confidence classes, and their properties will be published in a separate paper (C. Magagnoli et al., in preparation).

\citet{kartaltepemor15} define three mutually exclusive classes for potentially interacting and merging galaxies for the visual morphological classification scheme, which we will refer to here as Merger, Blended Interaction, and Non-blended Interaction. We apply further constraints on galaxies in these classes to select a sample of potential high confidence major interactions and mergers. The definitions of these classes and our further constraints are described below:

(i) \textbf{Merger}:  A galaxy that shows signs of a recent merger such as tidal tails, loops, double nuclei, or highly irregular outer isophotes is classified as a merger. We apply an additional constraint on the mass of the merged system to be greater than $1.25\times10^{10}$ M$_{\odot}$. If the minimum mass of the primary galaxy at a pre-merger stage is greater than 10$^{10}$ M$_{\odot}$ and the maximum mass ratio of the stellar mass of the primary to that of the secondary galaxy is 4 then the stellar mass of the merged galaxy system has to be greater than $1.25\times10^{10}$ M$_{\odot}$. We also require the redshift of the mergers to be between 0.5 and 3.0. Based on these criteria, we generated a sample of 66 high confidence major galaxy mergers. We show an example of a merger in the rightmost panel of Figure~\ref{fig:late_stage_mer}.

\begin{figure}[b]
    \centering
    \includegraphics[scale=0.34,trim = 1.3cm 1cm 0cm 1.3cm,clip=true]{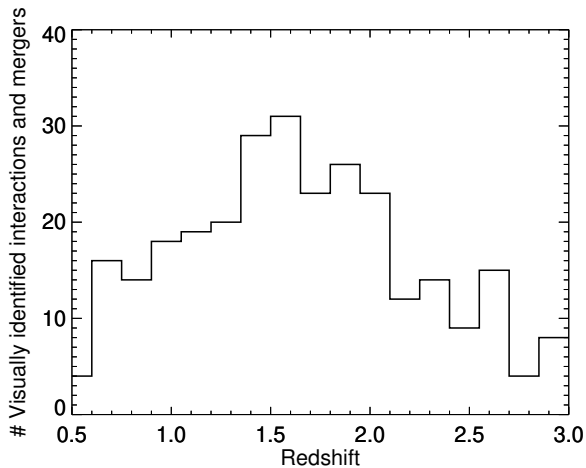}
    \caption{Photometric redshift distribution of the combined sample of visually identified high confidence mergers, blended interactions, and non-blended interactions. Note that this sample has a broader redshift distribution than the galaxy pair sample shown in Figure~\ref{fig:envzm_distr} with a median redshift of 1.6.}
    \label{fig:zdistr_late_stage}
\end{figure}

\begin{figure*}[t]
    \centering
    \hspace{-0.2in}
    \includegraphics[trim = 2cm 4.9cm 1.3cm 8.4cm,clip=true, scale=0.7]{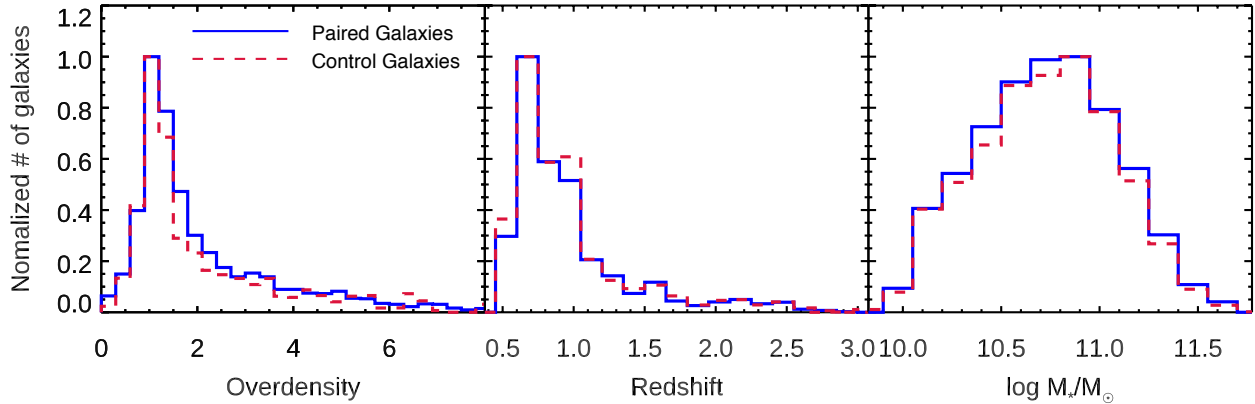}
    \caption{Environmental overdensity (left panel), spectroscopic redshift (middle panel), and stellar mass (right panel) distributions (normalized to the peak value) of 1345 spectroscopic galaxy pairs (solid blue line) (satisfying  $\Delta V<1000$\thinspace km s$^{-1}$, projected separation $<150$\thinspace kpc, mass ratio \thinspace $<4$, and spectroscopic redshift between 0.5 and 3) and the corresponding mass-, redshift-, and environment-matched control galaxies (red dashed line). }
    \label{fig:envzm_distr}
\end{figure*}

(ii) \textbf{Blended Interaction}: If a galaxy pair shows clear signs of tidal interactions (e.g., tidal arms, tidal bridges, dual asymmetries,  off-center isophotes, or other signs of morphological disturbance) and both galaxies are within the same \textit{H}-band segmentation map then the system is classified as a `Blended Interaction.' Classifiers choose this class over the merger class if two distinct galaxies are visible. In the case of more than one companion, the class is determined by the one that seems to dominate the morphology, which is typically the larger/brighter one. Since these sources are blended, the photometry corresponds to the combined system of the two galaxies, i.e., the properties of the system such as the stellar mass and photometric redshift correspond to the combined system. Hence, we apply the same additional constraint on the mass of the combined system as in the merger class, i.e., the stellar mass of the combined blended system has to be greater than $1.25\times10^{10}$\thinspace M$_{\odot}$. We also require the redshift value of the system to be $0.5<z<3.0$. We visually identify the photocenter of each of the galaxies and use the photometric redshift for the combined system to estimate the projected separation of the two galaxies. Using these constraints, we generated a sample of 100 high confidence galaxy pair systems going through a close interaction. The median projected separation for this sample is 7.73 \thinspace kpc. We show an example of a blended interaction in the middle panel of Figure~\ref{fig:late_stage_mer}.

(iii) \textbf{Non-blended Interaction}: The only difference between this class and the `Blended Interaction' class, is that in this class, the two interacting galaxies do not belong to the same \textit{H}-band segmentation map so both galaxies have their own measurements of the photometric properties. Hence, we apply constraints to both galaxies. The stellar mass of the secondary galaxy has to be greater than 10$^{10}$\thinspace M$_{\odot}$, the stellar mass ratio of the primary to secondary galaxy has to be less than four, and their photometric redshift error bars have to overlap with each other. Our sample of non-blended interactions consists of 61 galaxy pairs, i.e., 122 galaxies. The leftmost image in Figure~\ref{fig:late_stage_mer} shows an example of a non-blended interaction, showing two distinct galaxies in different segmentation maps with visible signs of interaction such as tidal tails. The median projected separation for this sample is 13.15 \thinspace kpc.

Figure~\ref{fig:zdistr_late_stage} shows the photometric redshift distribution of the combined sample of high confidence mergers, blended interactions, and non-blended interactions. The photometric redshift distribution of the visually identified mergers and interactions (median $z\sim1.6$) is much broader than the spectroscopic redshift distribution of the pair sample (median $z\sim1$).

\subsection{Control Samples}\label{sec:controls_def}

To isolate the effects of galaxy interactions on galaxy properties, the effects of other strongly variable properties affecting AGN activity like the stellar mass, redshift, and environment of the galaxy have to be controlled for. The distribution of these properties for the paired galaxies could be significantly different from the overall distribution of galaxies. Therefore, if we randomly select isolated galaxies, the distribution of their properties (such as mass, redshift, and environmental density) could be different from that of the pairs. We select a sample of isolated galaxies with similar stellar mass, redshift, and environment distributions as our paired galaxies. Since the spectroscopic completeness varies with each field and is highly correlated with properties such as stellar mass, star formation rate, and the presence of an AGN, we require our controls for the galaxy pair sample to have spectroscopic redshifts and to be selected from the same field. For the control sample for the visually identified and interactions and mergers, we do not require spectroscopic redshifts.

\begin{figure}[ht]
    \centering
    \hspace*{-0.2in}
    \includegraphics[scale=0.59, trim = 1.79cm 0.4cm 1.3cm 0.87cm,clip=true]{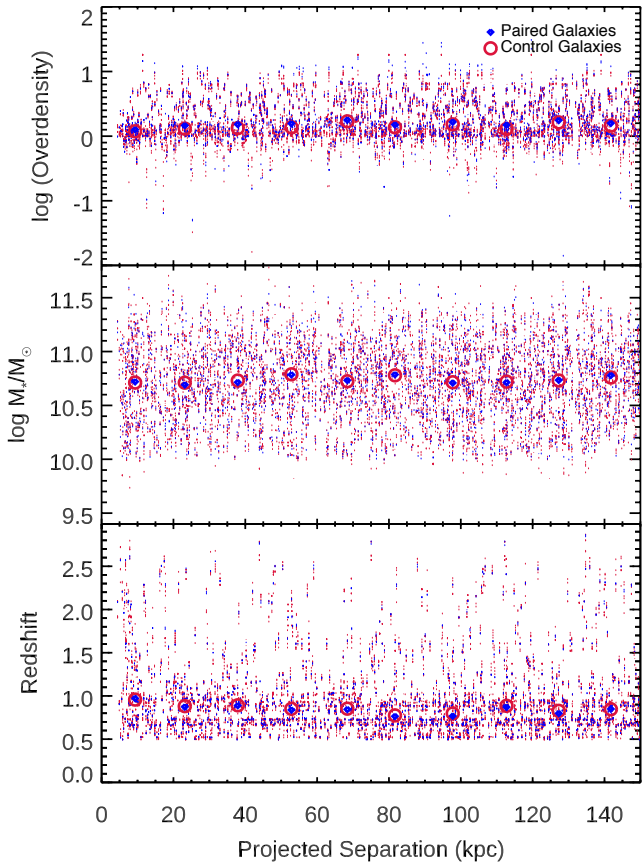}
    \caption{The small dots show the redshift (lower panel), stellar mass (middle panel), and overdensity (upper panel) values of individual paired (blue) and their corresponding control (red) galaxies as a function of the projected separation of the paired galaxies. For control galaxies, the projected separation value of the corresponding paired galaxy is used. The median properties of all the paired and control galaxies within projected separation bins of 10 \thinspace kpc width are shown by red diamond and blue open circle, respectively. While the paired galaxy sample spans a wide range of mass, redshift, and environmental overdensity, the median values of these properties do not vary significantly with the projected separation.
}
    \label{fig:distr_panel_wrt_sep}
\end{figure}

We create a parent sample of isolated galaxies with no major or minor companion (within a mass ratio of 10) within a $\Delta z$ corresponding to a relative velocity of less than 5000\thinspace km s$^{-1}$, out to a projected separation of 150\thinspace kpc. We also exclude the visually identified interactions and mergers described in the previous subsection from the control candidate samples. We then match the mass, redshift, and local galaxy environment of the controls with that of the paired galaxies. The environmental overdensity (ratio of the density around the position and redshift of the galaxy to that of the median density in that redshift bin) for galaxies in the COSMOS field was estimated using redshift-dependent `weighted' adaptive kernel density maps generated by \citet{darvish2015}. For the CANDELS fields, the density estimation was carried out using the Voronoi Tessellation method described by \citet{lemaux2017} and \citet{tomczak2017}. Though these methods are slightly different, previous work has shown that the results are consistent with one another \citep{darvish2015}, and we find no significant systematic differences. In both cases, we calculated the overdensity from the density measurements in a consistent way.

To generate the final control sample, for each galaxy in our galaxy pair sample, we select three control galaxies from the above mentioned control parent sample by minimizing $ (\Delta\log{M_*})^2+(\Delta z)^2+(1/40)(\Delta overdensity)^2$. Considering the range and distribution of overdensity, redshift, and stellar mass, we used a weighing factor of 1/40 for the overdensity to obtain the best match in all three dimensions so that the overdensity-matching does not dominate. For more than 90\% of paired galaxies, the controls match within a stellar mass of 0.15 \thinspace dex, spectroscopic redshift within 0.15, and overdensity within 1. Our final control sample contains 8070 (6399) control galaxies for pairs with $\Delta V<1000$\thinspace (500)\thinspace km s$^{-1}$, out of which 8034 (6374) galaxies are unique.

The normalized environmental overdensity, redshift, and stellar mass distribution of the final galaxy pair sample and corresponding control galaxy sample is shown in Figure~\ref{fig:envzm_distr}. The distribution of these quantities as a function of the projected separation is shown in Figure~\ref{fig:distr_panel_wrt_sep}. These plots show that the galaxy pairs and controls have similar environmental overdensity, redshift, and stellar mass distributions, crucial for our analysis. The middle panel in Figure~\ref{fig:envzm_distr} shows that the number of paired galaxies increases with redshift out to $z\sim0.8$ and then decreases, with a median value of 1.0. The right panel shows that the sample is mostly uniform for masses between about  10$^{10}$ \thinspace M$_{\odot}$ and 10$^{11}$ \thinspace M$_{\odot}$, after which it rapidly decreases for increasing mass, with very few galaxies above 10$^{11.5}$ \thinspace M$_{\odot}$. Figure~\ref{fig:distr_panel_wrt_sep} shows that while the paired galaxy sample spans a wide range of mass, redshift, and environmental overdensity, the median value of these properties do not vary significantly with the projected separation.

\section{Analysis of AGN Activity}\label{sec:agn-enh}

In this section, we discuss the identification of AGN in the X-ray and IR, and measurement of the AGN fraction for the spectroscopic paired galaxies, visually identified mergers and interactions, and control galaxies. We then estimate the level of AGN enhancement and its dependence on the projected separation of galaxy interactions.

\subsection{AGN Identification}\label{sec:agnidentification}

\subsubsection{X-ray AGN} \label{sec:xray_agn_def}
We use the Chandra X-ray observations (Section \ref{subsec:xraydata}) to identify X-ray selected AGN. For the X-ray sources among the spectroscopic pairs and their corresponding control samples, we computed the total X-ray luminosity ${L_{X}}$ following the method of Marchesi et al. 2016, using the spectroscopic redshift $z$) and X-ray flux ($F_{X}$) values in

\begin{equation}
   {L_{X}} = F_{X} \times 4 \pi d^2 \times k(z),
\end{equation}

\noindent where

\begin{equation}
     {k(z)} = (1 + {z})^{(\Gamma - 2)},
\end{equation}

\noindent d is the luminosity distance for a given redshift, \textit{k(z)} is the \textit{k}-correction, and $\Gamma$ = 1.4 is the slope of the power law \citep{marchesi2016}. We identify the sources with the total (full band: $0.5 - 10$\thinspace keV) X-ray luminosity of greater than 10$^{42}$\thinspace erg s$^{-1}$ as X-ray AGN \citep[e.g.,][]{moran1999}.  This luminosity cut ensures that the observed flux is almost completely dominated by the AGN and the contamination due to star formation is negligible. Although this requirement may miss many low-luminosity and/or highly dust-obscured AGN, in comparison with other selection methods (e.g. optical, IR, radio), X-ray identification of AGN provides a clean AGN sample.

\subsubsection{IR AGN}
We use the \textit{Spitzer}/IRAC observations described in Section \ref{subsec:irdata} to identify IR AGN using two different sets of selection criteria \citep{stern2005,donley2012}. While the \citet{stern2005} criteria select a more complete sample of AGN, this sample is also subject to a large amount of contamination from star formation, while the \citet{donley2012} selected sample is less contaminated but also less complete. We include both samples in our analysis for comparison.

Galaxies with dominant AGN emission usually follow a characteristic red power law in the IR \citep[f$_{\nu}\propto\nu^{\alpha}$ with $\alpha$ $\le$ -0.5;][]{alonso2006}. Therefore, IR power law selection can be used to select a clean AGN sample. The \citet{donley2012} criteria provide reliable identification of luminous AGN with minimal contamination from star formation. To satisfy the \citet{donley2012} criteria, objects must be detected in all four IRAC bands, and their colors lie within the following IRAC color-color region:

\begin{equation}
   \textit{x} = \log_{10}\left(\frac{\textit{f}_{5.8\mu m}}{\textit{f}_{3.6\mu m}} \right),       \textit{y} = \log_{10}\left(\frac{\textit{f}_{8.0\mu m}}{\textit{f}_{4.5\mu m}} \right),
\end{equation}
\begin{equation}
   \textit{x} \geq 0.08 \: \wedge \:  \textit{y} \geq 0.15, \:
   \end{equation}
   \begin{equation}
  \textit{y} \geq (1.21 \times \textit{x}) - 0.27 \: \wedge \:
    \textit{y} \leq (1.21 \times \textit{x}) + 0.27,
     \end{equation}
    \begin{equation}
    \textit{f}_{4.5\mu m} > \textit{f}_{3.6\mu m} \wedge \:
    \textit{f}_{5.8\mu m} > \textit{f}_{4.5\mu m} \wedge \:
    \textit{f}_{8.0\mu m} > \textit{f}_{5.8\mu m},
\end{equation}

\noindent where $f_\lambda$ is the flux of the galaxy at wavelength $\lambda$, and $`\wedge$' is the logical ``AND'' operator. Using these criteria, we identify 31 Donley IR AGN in the paired galaxy sample, and 99 AGN in the control galaxy sample. The combined sample of visually identified mergers and interactions contains 5 Donley IR AGN, and their control sample contains 19 AGN.

The \citet{stern2005} IRAC color-color selection criteria used to identify IR AGN is defined as

\begin{equation}
    ([5.8] - [8.0]) > 0.6 ,
\end{equation}
\begin{equation}
 ([3.6] -[4.5]) > 0.2 \times ([5.8] - [8.0]) + 0.18 ,
\end{equation}
\begin{equation}
([3.6] - [4.5]) > 2.5 \times ([5.8] - [8.0]) - 3.5,
\end{equation}

\noindent where [$\lambda$] is the Vega Magnitude of the galaxy at wavelength $\lambda$ in \thinspace$\mu$m. Using these criteria, we identify 106 Stern IR AGN in paired galaxies and 296 in control galaxies. The combined sample of visually identified merger and interaction has 47, and their combined control sample has 129 Stern IR AGN.

There are six paired and 35 corresponding control galaxies that have both X-ray and IR AGN (using the \citet{donley2012} criteria). There are too few objects in this overlapping sample to allow us to analyze how the fraction of these relates to other properties, such as redshift or pair separation.  There are four galaxies in the visually identified interaction and merger sample and eight corresponding control galaxies that have both X-ray and IR AGN. In total, there are 194 paired galaxies and 584 control galaxies that have either X-ray or IR AGN. Likewise, there are 28 galaxies in the visually identified interaction and merger sample and 78 galaxies in the corresponding control samples have either X-ray or IR AGN.

\subsection{AGN Enhancement in Spectroscopic Galaxy Pairs}\label{sec:agn-enh-def}
To estimate the level of AGN enhancement in our galaxy pair sample relative to the control galaxies, we divide the sample of galaxy pairs into projected separation bins (depending on the number of AGN in a given bin) with a width of 25 \thinspace kpc (6 bins) or 50\thinspace kpc (3 bins). We define the X-ray (or IR) AGN fraction as the ratio of the total number of galaxies having an X-ray (or IR) AGN to that of the total number of galaxies, i.e,

\begin{equation}
{AGN Fraction}=\frac{\#\:AGN}{\#\:Total},
\end{equation}

\noindent where AGN Fraction is for paired (control) galaxies within a given projected separation bin,  \textit{\# AGN} is the number of paired (control) galaxies with an AGN in the given projected separation bin, and \textit{\# Total} is the total number of paired (control) galaxies in the given projected separation bin. For each separation bin, we calculate the AGN fraction in the paired galaxy sample and the corresponding control galaxy sample.

\begin{figure*}[t]
    \centering
    \includegraphics[scale=0.83,angle=0, trim = 2.7cm 2.4cm 3.78cm 11cm,clip=true]{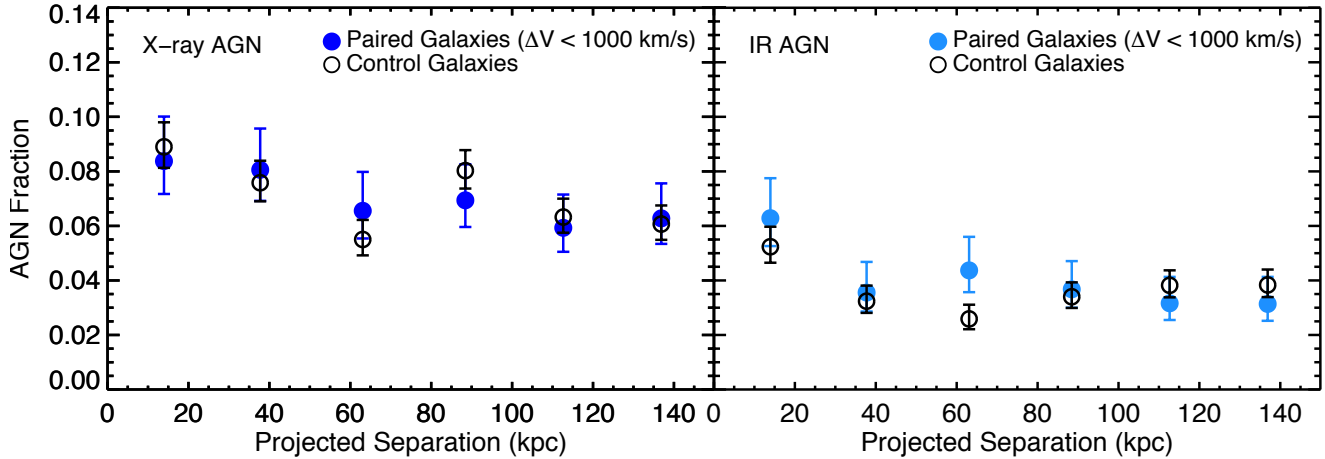}
    \caption{(Left: X-ray, Right: IR)  AGN fraction (defined by the ratio of the number of galaxies with an AGN to that of the total number of galaxies in a given projected separation bin). The paired galaxies ($\Delta V<1000$\thinspace km s$^{-1}$) are indicated by dark blue filled circles, and light blue filled circles, respectively. The black open circles in both panels show the corresponding mass-, redshift- and environment-matched control galaxies. The error bars on each point reflect the 1$\sigma$ binomial confidence limits, following the method of \citet{cameron2011}. IR AGN are identified using \citet{stern2005} criteria. In both panels, the AGN fraction in paired galaxies slightly increases with decreasing separation. However, the AGN fraction of the control sample also increases.}
    \label{fig:frac_comb}
\end{figure*}

For the $\Delta V<1000$\thinspace km s$^{-1}$ kinematic pair sample, the left panel of Figure \ref{fig:frac_comb} shows the X-ray AGN fraction for six different projected separation bins of width 25 \thinspace kpc each. While there is a slight increase in the AGN fraction of the paired galaxies with decreasing separation (with a value of $8.4^{+1.6}_{-1.2}\%$ at $<25$\thinspace kpc), the AGN fraction of the control sample also slightly increases. The right panel of Figure \ref{fig:frac_comb} shows the IR AGN fraction using the \cite{stern2005} selection criteria for the same six projected separation bins. Just as for the X-ray AGN fraction, the IR AGN fraction of paired galaxies increases with decreasing projected separation, with a value of $6.3^{+1.47}_{-1.02}\%$ at $<25$\thinspace kpc. However, the AGN fraction of the controls also increases in a similar manner. For all bins, the AGN fraction of pairs and controls are similar to each other.

We then define the AGN enhancement as the ratio of the AGN fraction of paired galaxies to that of the corresponding control galaxies, i.e,

\begin{multline}
{AGN Enhancement} = \frac{AGN\:Fraction_{Pairs}}{AGN\:Fraction_{Controls}},
\end{multline}

\noindent where AGN Fraction$_{Pairs}$ and AGN Fraction$_{Controls}$ are the AGN fraction values of the paired and control galaxy samples, respectively, in a given projected separation bin. We assume binomial statistics to calculate 1$\sigma$ errors \citep{cameron2011} in the AGN fraction, and then propagate them to compute the errors in AGN enhancement. Throughout this work, we carry out a separate analysis for X-ray and IR AGN enhancements.

  \begin{figure*}[t]
    \centering
    \includegraphics[scale=0.83,angle=0, trim = 2.7cm 2.4cm 3.78cm 11cm,clip=true]{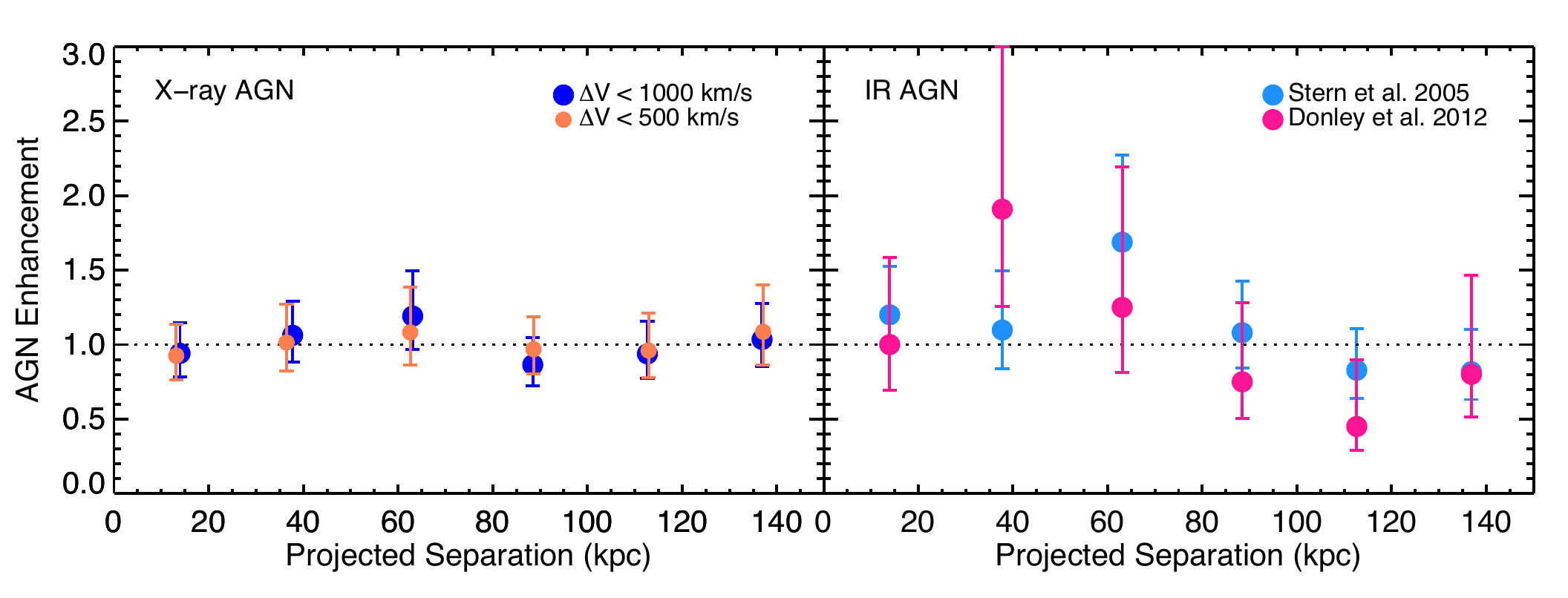}
   \caption{The level of (left: X-ray, right: IR) AGN enhancement (defined by the ratio of the AGN fraction of paired galaxies to that of the corresponding control galaxies) as a function of the projected separation of the paired galaxies. The error bars on each point reflect the $1\thinspace\sigma$ binomial confidence limits, following the method of \citet{cameron2011}. The horizontal dashed line corresponds to an AGN enhancement value of one, i.e., the AGN fraction of the paired galaxy sample is the same as the AGN fraction of the corresponding control sample and therefore signify an absence of interaction-induced AGN enhancement. Left panel: The dark blue filled circles and orange filled smaller circles correspond to the spectroscopic galaxy pairs with $\Delta V<1000$\thinspace km s$^{-1}$ and $\Delta V<500$\thinspace km s$^{-1}$, respectively. Right panel: The IR AGN identification is based on the selection criteria of \citet{stern2005} (light blue filled circle) and \citet{donley2012} (deep pink filled circles) applied to the IRAC observations of paired ($\Delta V<1000$\thinspace km s$^{-1}$) and control galaxies. The X-ray and IR enhancement values for the paired galaxy sample with $\Delta V<1000$\thinspace km s$^{-1}$ are provided in Table~\ref{table:enh_alll_allz} and Table~\ref{table:enh_ir}, respectively.
}
    \label{fig:enh_comb}
\end{figure*}

We calculate the X-ray AGN enhancement for spectroscopic galaxy pairs and present the results in the left panel of Figure~\ref{fig:enh_comb} and Table~\ref{table:enh_alll_allz} for both the $\Delta V<1000$\thinspace km s$^{-1}$ and the $\Delta V<500$\thinspace km s$^{-1}$ samples. The horizontal dashed line corresponds to an AGN enhancement value of one, i.e., the AGN fraction of the paired galaxy sample is the same as the AGN fraction of its control sample, therefore, indicating an absence of enhancement. We find an AGN enhancement of 0.94$^{+0.21}_{-0.16}$ for the closest separation bin for pairs with $\Delta V<1000$\thinspace km s$^{-1}$. We do not find a statistically significant enhancement at any separation for any of the velocity cuts used. The results of both samples are consistent with each other, which could be due to the fact that galaxies with $\Delta V<500$\thinspace km s$^{-1}$ dominate the $\Delta V<1000$\thinspace km s$^{-1}$ sample. Table~\ref{table:enh_alll_allz} presents the values of the number of paired and their corresponding control galaxies, the number of X-ray AGN and AGN fraction in these samples, and the corresponding X-ray AGN enhancement in the paired galaxies used for Figure \ref{fig:enh_comb}. These values include the full sample of X-ray AGN at all luminosities across the complete redshift range of $0.5<z<3$ with $\Delta V<1000$\thinspace km s$^{-1}$.

The right panel of Figure \ref{fig:enh_comb} shows the level of IR AGN enhancement in the  $\Delta V<1000$\thinspace km s$^{-1}$ kinematic pair sample at $0.5<z<3.0$ using both the \citet{stern2005} and \citet{donley2012} criteria. Since the $\Delta V<500$\thinspace km s$^{-1}$ sample is significantly smaller with a limited number of Donley IR AGN, we do not include it here. At the smallest separation, we calculate the Donley IR AGN enhancement to be 1.00$^{+0.58}_{-0.31}$ and the Stern IR AGN enhancement to be 1.06$^{+0.38}_{-0.26}$, consistent within error bars.  Table~\ref{table:enh_ir} includes the values used for the \citet{donley2012} criteria identified AGN enhancement.  We do not find a statistically significant enhancement for IR AGN in any separation bin. In the figure, the error bars for the Stern IR AGN are smaller than the error bars for the Donley IR AGN since the \citet{stern2005} criteria identify a larger number of AGN than the \citet{donley2012} criteria. We also tested the effect of applying different $S/N$ cuts to the IRAC fluxes and do not find a significant difference when using $S/N>3$ or $S/N>5$ cut.

We find a similar result (no significant enhancement) when considering the combined X-ray and IR AGN sample. There are 194 paired galaxies in this category, i.e., pairs in which at least one galaxy contains either X-ray or IR AGN. Furthermore, six paired galaxies have both, an X-ray and IR-selected AGN, but there are too few AGN to be further divided into bins for analysis.

\begin{deluxetable*}{lcccccc}

\tablewidth{0pt}
\tablecaption{X-ray AGN Enhancement: All Fields ($Lx>10^{42}$\thinspace erg s$^{-1}$, $0.5<z<3.0$, $\Delta V<1000$\thinspace km s$^{-1}$)}

\tablehead{\colhead{} & \colhead{$0<d<25$} & \colhead{$25<d<50$} & \colhead{$50<d<75$} & \colhead{$75<d<100$} & \colhead{$100<d<125$} & \colhead{$125<d<150$}}
\startdata
 Paired Galaxies & 382 & 422 & 412 & 490 & 506 & 478\\
 AGN & 32 & 34 & 27 & 34 & 30 & 30\\
AGN Fraction (\%) & 8.4$^{+1.6}_{-1.2}$ & 8.1$^{+1.5}_{-1.1}$ & 6.6$^{+1.4}_{-1.0}$ & 6.9$^{+1.3}_{-0.9}$ & 5.9$^{+1.2}_{-0.8}$ & 6.3$^{+1.3}_{-0.9}$\\ \hline
Control Galaxies & 1146 & 1266 & 1236 & 1470 & 1518 & 1434\\
AGN & 102 & 96 & 68 & 118 & 96 & 87\\
AGN Fraction (\%) & 8.9$^{+0.9}_{-0.8}$ & 7.6$^{+0.8}_{-0.7}$ & 5.5$^{+0.7}_{-0.6}$ & 8.0$^{+0.8}_{-0.7}$ & 6.3$^{+0.7}_{-0.6}$ & 6.1$^{+0.7}_{-0.6}$ \\ \hline
AGN Enhancement & 0.94$^{+0.21}_{-0.16}$ & 1.06$^{+0.23}_{-0.18}$ & 1.19$^{+0.30}_{-0.22}$ & 0.86$^{+0.18}_{-0.14}$ & 0.94$^{+0.22}_{-0.16}$ & 1.03$^{+0.24}_{-0.18}$\\
\enddata
\tablecomments{The projected separation($d$) is measured in kpc.}
\label{table:enh_alll_allz}
\end{deluxetable*}

\begin{deluxetable*}{lcccccc}

\tablewidth{0pt}
\tablecaption{IR AGN Enhancement: All Fields (\citet{donley2012} criteria, $0.5<z<3.0$, $\Delta V<1000$\thinspace km s$^{-1}$)}

\tablehead{\colhead{} & \colhead{$0<d<25$} & \colhead{$25<d<50$} & \colhead{$50<d<75$} & \colhead{$75<d<100$} & \colhead{$100<d<125$} & \colhead{$125<d<150$}}
\startdata
 Paired Galaxies & 382 & 422 & 412 & 490 & 506 & 478\\
 AGN & 7 & 7 & 5 & 5 & 3 & 4 \\
AGN Fraction (\%) & 1.8$^{+0.9}_{-0.5}$ & 1.7$^{+0.9}_{-0.4}$ & 1.2$^{+0.8}_{-0.3}$ & 1.0$^{+0.7}_{-0.3}$ & 0.6$^{+0.6}_{-0.2}$ & 0.8$^{+0.6}_{-0.3}$ \\ \hline
 Control Galaxies & 1146 & 1266 & 1236 & 1470 & 1518 & 1434\\
 AGN & 21 & 11 & 12 & 20 & 20 & 15\\
AGN Fraction (\%) & 1.8$^{+0.5}_{-0.3}$ & 0.9$^{+0.3}_{-0.2}$ & 0.9$^{+0.4}_{-0.2}$ & 1.4$^{+0.4}_{-0.2}$ & 1.3$^{+0.4}_{-0.2}$ & 1.0$^{+0.3}_{-0.2}$ \\ \hline
AGN Enhancement & 1.00$^{+0.58}_{-0.31}$ & 1.90$^{+1.25}_{-0.65}$ & 1.25$^{+0.94}_{-0.44}$ & 0.75$^{+0.53}_{-0.25}$ & 0.45$^{+0.45}_{-0.16}$ & 0.80$^{+0.66}_{-0.28}$\\
\enddata
\tablecomments{The projected separation($d$) is measured in kpc.}
\label{table:enh_ir}
\end{deluxetable*}

\begin{deluxetable}{cccc}

\tablewidth{0pt}
\tablecaption{X-ray Luminosity-Redshift ($L_{\rm{X}}$-z) Bins Used for Analysis}
\tablehead{\colhead{Panel} & \colhead{$\log \thinspace(L_{\rm{X}}\thinspace(\rm{erg \thinspace s}^{-1}))$} & \colhead{Redshift (z)} & \colhead{Field(s)}}
\startdata
Low $L_{\rm{X}}$ & $42.0<\log(L_{\rm{X}})<43.2$ & $0.5<z<2.0$ & GOODS\\
Moderate $L_{\rm{X}}$ & $43.2<\log(L_{\rm{X}})<43.7$ & $0.5< z\thinspace<2.0$ & All\\
High $L_{\rm{X}}$ & $43.7<\log(L_{\rm{X}})$ & $0.5<z<3.0$ & All
\enddata
\tablecomments{L$_{\rm{X}}$ denotes the full band ($0.5-10$\thinspace keV) X-ray luminosity of a galaxy in erg s$^{-1}$.}
\label{table:lz_cut}
\end{deluxetable}

\begin{figure}[h]
    \centering
    \vspace*{-0.3in}
    \hspace*{-0.3in}
    \includegraphics[scale=0.39,trim = 1.4cm 1.3cm 1.4cm 1.3cm,clip=true]{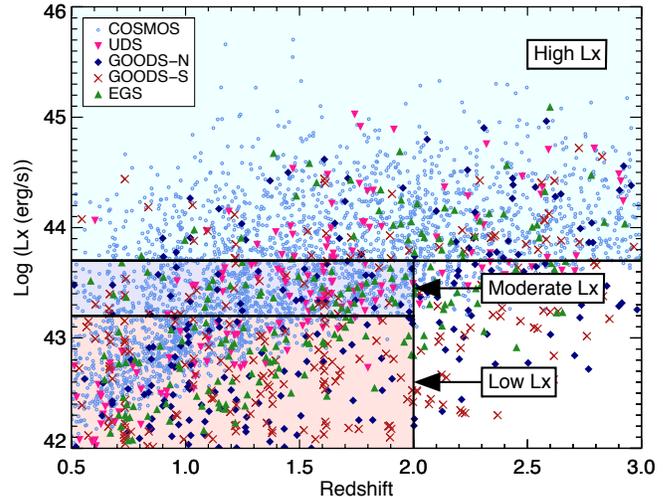}
    \caption{The distribution of the total, i.e., full band (0.5\thinspace keV\thinspace--\thinspace10\thinspace keV) X-ray luminosity ($L_{\rm{X}}$) with respect to redshift for all X-ray AGN ($L_{\rm{X}}>10^{42}$\thinspace erg s$^{-1}$) in the COSMOS and CANDELS fields. In the plot, the pink downward triangles, navy diamonds, maroon crosses, green upward triangles, and small light blue circles correspond to all X-ray AGN in UDS, GOODS-N, GOODS-S, EGS, and COSMOS, respectively. Highlighted are the three $L_{\rm{X}}$-z bins used in our analysis. The light red shaded region (Low $L_{\rm{X}}$ bin) X-ray sources with $42.0<\log(L_{\rm{X}})<43.2$ at $0.5<z<2.0$ in the GOODS fields. The lavender (Moderate $L_{\rm{X}}$: $43.2<\log(L_{\rm{X}})<43.7$) and light blue shaded (High $L_{\rm{X}}$: $43.7<\log(L_{\rm{X}})$) regions correspond to sources in all the fields with $0<z<2$ and $0<z<3$, respectively.}

    \label{fig:agn_lum_z}
\end{figure}

\begin{figure}[ht]
    \centering
    \hspace*{-0.35in}
    \includegraphics[scale=0.6,trim = 1cm 1.3cm 8cm 1.0cm,clip=true]{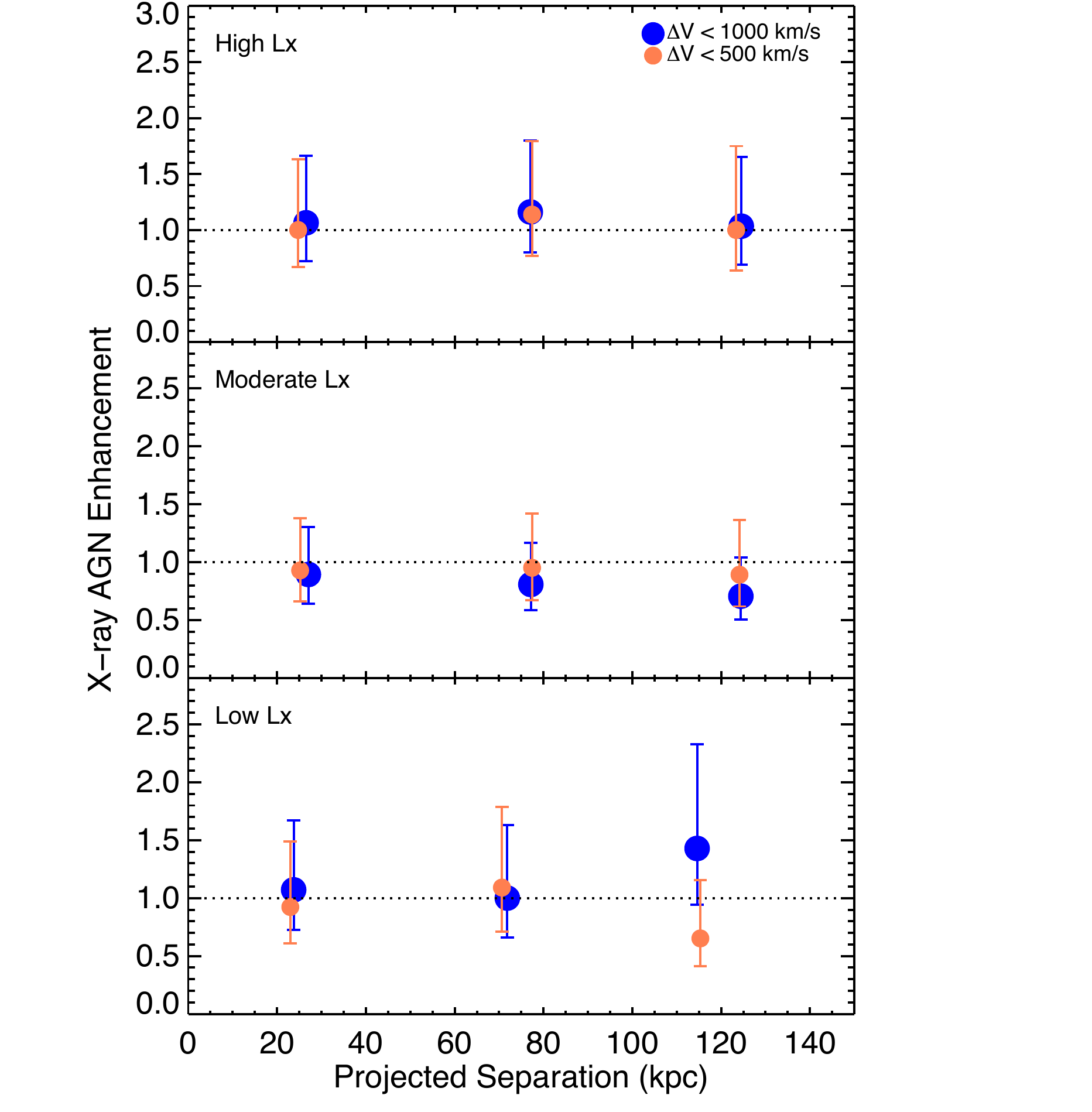}
    \caption{The X-ray AGN enhancement as a function of the projected separation of the paired galaxies with $\Delta V<1000$\thinspace km s$^{-1}$ (large filled blue circles) and $\Delta V<500$\thinspace km s$^{-1}$ (small filled orange circles), split into three different $L_{\rm{X}}$-z bins. The lower panel (Low $L_{\rm{X}}$ bin) corresponds to the galaxies in the GOODS-North and GOODS-South fields with $0.5<z<2.0$ and $42.0<\log(L_{\rm{X}})<43.2$. The middle panel (Moderate $L_{\rm{X}}$ bin) corresponds to the galaxies in all fields (CANDELS and the full COSMOS field) with $0.5<z<2.0$ and $43.2<\log(L_{\rm{X}})<43.7$. The upper panel (High $L_{\rm{X}}$ bin) corresponds to galaxies in all the fields with $0.5<z<3.0$ and $43.2<\log(L_{\rm{X}})<43.7$. The values of the luminosity cut at a given redshift are chosen based on X-ray completeness. The symbols for the pair sample match those in the left panel of Figure~\ref{fig:enh_comb}. The $L_{\rm{X}}$-z bins are defined in Table~\ref{table:lz_cut} and illustrated in Figure \ref{fig:agn_lum_z}.
}
    \label{fig:enh_ladder}
\end{figure}

 The depth (and therefore the sensitivity) of the Chandra X-ray observations varies over the CANDELS and COSMOS fields. Figure \ref{fig:agn_lum_z} shows the total (0.5\thinspace keV -- 10\thinspace keV) X-ray luminosity ($L_{\rm{X}}$)  distribution as a function of redshift for all X-ray AGN in all fields, highlighting that the GOODS fields have the deepest and the COSMOS field has the shallowest X-ray observations. Since our galaxy pair and control samples consist of galaxies from all of the above-mentioned fields and we want to compare similar types of AGN across different fields at different redshifts, it is necessary to be consistent and use the same constraints to select AGN with similar luminosities from all the fields.

Considering the variation in  X-ray completeness for the different fields, we apply three different luminosity-redshift ($L_{\rm{X}}$-z) cuts as defined in Table~\ref{table:lz_cut} and Figure~\ref{fig:agn_lum_z} to identify X-ray selected AGN in paired and control galaxies: (i) Low $L_{\rm{X}}$ AGN: $42<\log(L_{\rm{X}} (erg/s))<43.2$ and $0.5<z<2.0$ for the GOODS (North and South) fields, (ii) Moderate $L_{\rm{X}}$ AGN: $43.2<\log(L_{\rm{X}} (erg/s))<43.7$ and $0.5<z<2.0$ for all fields, (iii) High $L_{\rm{X}}$ AGN: $\log(L_{\rm{X}} (erg/s))>43.7$ and $0.5>z<3.0$ for all fields, corresponding to high luminosity AGN and dominated by quasars $(\log(L_{\rm{X}})>44)$.

The X-ray AGN enhancement for these X-ray complete $L_{\rm{X}}$-z cut bins in the $\Delta V<1000$ \thinspace km s$^{-1}$ and $\Delta V<500$ \thinspace km s$^{-1}$ pairs samples are shown in Figure \ref{fig:enh_ladder}. The lower, middle, and upper panels correspond to the Low $L_{\rm{X}}$, Moderate $L_{\rm{X}}$, and High $L_{\rm{X}}$ bins, respectively. The X-ray AGN enhancement results for the $\Delta V<1000$ \thinspace km s$^{-1}$ pair sample are presented in Table ~\ref{table:enh_ladder}. We do not see any significant enhancement in any of the three luminosity bins at any separation. The results do not change significantly if we use a stricter cut on the relative velocity difference ($\Delta V<500$ \thinspace km s$^{-1}$) as shown in the figure. The $\Delta V<1000$ \thinspace km s$^{-1}$ value is sightly elevated for the largest separation bin at low Lx, however, the $\Delta V<500$ \thinspace km s$^{-1}$ value shows the opposite. The deviation of these enhancement values from a value of one (no enhancement) is not statistically significant due to the small number of AGN in these bins.

\begin{deluxetable*}{l|ccc|ccc|ccc}
\tablewidth{0pt}
\tablecaption{X-ray AGN Enhancement in $\Delta V<1000$\thinspace km s$^{-1}$ Sample in Different $L_{\rm{X}}$-z bins: Figure~\ref{fig:enh_ladder}}
\tablehead{\colhead{} & \colhead{} & \colhead{Low $L_{\rm{X}}$} & \colhead{} & \colhead{} & \colhead{Moderate $L_{\rm{X}}$} & \colhead{} & \colhead{} & \colhead{High $L_{\rm{X}}$} & \colhead{}\\
\colhead{} & \colhead{d[0,50]} & \colhead{d[50,100]} & \colhead{d[100,150]} & \colhead{d[0,50]} & \colhead{d[50,100]} & \colhead{d[100,150]} & \colhead{d[0,50]} & \colhead{d[50,100]} & \colhead{d[100,150]}}
\startdata
Paired Galaxies & 116 & 98 & 116 & 742 & 876 & 926 & 804 & 902 & 984\\
 AGN & 10 & 8 & 10 & 14 & 14 & 12 & 11 & 12 & 10\\
AGN Fraction (\%) & 8.6$^{+3.3}_{-1.9}$ & 8.2$^{+3.6}_{-1.9}$ & 8.6$^{+3.3}_{-1.9}$ & 1.9$^{+0.6}_{-0.4}$ & 1.6$^{+0.5}_{-0.3}$ & 1.3$^{+0.5}_{-0.3}$ & 1.4$^{+0.5}_{-0.3}$ & 1.3$^{+0.5}_{-0.3}$ & 1.0$^{+0.4}_{-0.2}$\\\hline
 Control Galaxies & 348 & 294 & 348 & 2226 & 2628 & 2778 & 2412 & 2706 & 2952\\
 AGN & 28 & 24 & 21 & 47 & 52 & 51 & 31 & 31 & 29\\
AGN Fraction (\%) & 8.0$^{+1.7}_{-1.2}$ & 8.1$^{+1.9}_{-1.3}$ & 6.0$^{+1.5}_{-1.0}$ & 2.1$^{+0.3}_{-0.3}$ & 2.0$^{+0.3}_{-0.2}$ & 1.8$^{+0.3}_{-0.2}$ & 1.3$^{+0.3}_{-0.2}$ & 1.1$^{+0.2}_{-0.2}$ & 1.0$^{+0.2}_{-0.2}$\\\hline
AGN Enhancement & 1.07$^{+0.60}_{-0.35}$ & 1.00$^{+0.63}_{-0.34}$ & 1.43$^{+0.90}_{-0.48}$ & 0.89$^{+0.41}_{-0.25}$ & 0.81$^{+0.36}_{-0.22}$ & 0.71$^{+0.33}_{-0.20}$ & 1.06$^{+0.60}_{-0.34}$ & 1.16$^{+0.63}_{-0.36}$ & 1.03$^{+0.62}_{-0.34}$\\
\enddata
\tablecomments{d[x,y]: x\thinspace$<$\thinspace Projected Separation (d\thinspace/\thinspace kpc)\thinspace$<$\thinspace y.}
\label{table:enh_ladder}
\end{deluxetable*}

\begin{deluxetable*}{l|ccc|ccc|ccc}
\tablewidth{0pt}
\tablecaption{X-ray AGN Enhancement for Visually Identified Mergers and Interactions: Figure~\ref{Fig:enh_mor_mer_xray}}
\tablehead{\colhead{} & \multicolumn{3}{c}{Meger} & \multicolumn{3}{c}{Blended Int} & \multicolumn{3}{c}{Non-blended Int}\\
\colhead{} & \colhead{z[0.5,3.0]} & \colhead{z[0.5,1.6]} & \colhead{z[1.6,3.0]} & \colhead{z[0.5,3.0]} & \colhead{z[0.5,1.6]} & \colhead{z[1.6,3.0]} & \colhead{z[0.5,3.0]} & \colhead{z[0.5,1.6]} & \colhead{z[1.6,3.0]}}
\startdata
 Galaxies & 66 & 35 & 31 & 99 & 46 & 53 & 121 & 59 & 62\\
 AGN & 6 & 3 & 3 & 7 & 4 & 3 & 4 & 2 & 2\\
AGN Fraction (\%) & 9.1$^{+4.8}_{-2.4}$& 8.6$^{+7.2}_{-2.7}$ & 9.7$^{+7.9}_{-3.3}$ & 7.1$^{+3.5}_{-1.8}$ & 8.7$^{+6.0}_{-2.6}$ & 5.6$^{+4.9}_{-1.7}$ & 3.3$^{+2.5}_{-1.0}$ & 3.4$^{+4.1}_{-1.0}$ & 3.2$^{+4.0}_{-1.0}$\\ \hline
 Control Galaxies & 198 & 105 & 93 & 297 & 138 & 159 & 363 & 177 & 186\\
 AGN & 10 & 4 & 6 & 21 & 7 & 14 & 24 & 10 & 14\\
AGN Fraction (\%) & 5.1$^{+2.0}_{-1.1}$ & 3.8$^{+2.8}_{-1.1}$ & 6.5$^{+3.5}_{-1.7}$ & 7.1$^{+1.8}_{-1.2}$ & 5.1$^{+2.6}_{-1.3}$ & 8.8$^{+2.8}_{-1.8}$ & 6.6$^{+1.5}_{-1.1}$ & 5.6$^{+2.3}_{-1.3}$ & 7.5$^{+2.4}_{-1.5}$\\ \hline
AGN Enhancement & 1.8$^{+1.19}_{-0.63}$ & 2.2$^{+2.52}_{-0.96}$ & 1.5$^{+1.48}_{-0.62}$ & 1.0$^{+0.55}_{-0.30}$ & 1.7$^{+1.46}_{-0.67}$ & 0.64$^{+0.60}_{-0.24}$ & 0.5$^{+0.39}_{-0.17}$ & 0.6$^{+0.77}_{-0.23}$ & 0.43$^{+0.54}_{-0.16}$\\
\enddata
\tablecomments{z[a,b]: a\thinspace$<$\thinspace Redshift (z)\thinspace $<$\thinspace b.}
\label{table:enh_mor_mer_xray}
\end{deluxetable*}

\begin{deluxetable*}{l|ccc|ccc|ccc}
\tablewidth{0pt}
\tablecaption{\citet{stern2005} Identified IR AGN Enhancement for Visually Identified Mergers and Interactions: Figure~\ref{Fig:enh_mor_mer_ir}}
\tablehead{\colhead{} & \multicolumn{3}{c}{Meger} & \multicolumn{3}{c}{Blended Int} & \multicolumn{3}{c}{Non-blended Int}\\
\colhead{} & \colhead{z[0.5,3.0]} & \colhead{z[0.5,1.6]} & \colhead{z[1.6,3.0]} & \colhead{z[0.5,3.0]} & \colhead{z[0.5,1.6]} & \colhead{z[1.6,3.0]} & \colhead{z[0.5,3.0]} & \colhead{z[0.5,1.6]} & \colhead{z[1.6,3.0]}}
\startdata
 Galaxies & 66 & 35 & 31 & 99 & 46 & 53 & 121 & 59 & 62\\
 AGN & 12 & 5 & 7 & 13 & 5 & 8 & 22 & 13 & 9\\
AGN Fraction (\%) & 18.18$^{+5.6}_{-3.8}$ &  14.3$^{+7.9}_{-4.0}$ &  22.6$^{+9.0}_{-5.7}$ &  13.1$^{+4.1}_{-2.7}$ &  10.9$^{+6.3}_{-3.0}$ &  15.1$^{+6.2}_{-3.6}$ &  18.2$^{+4.0}_{-3.0}$ &  22.0$^{+6.2}_{-4.4}$ &  14.5$^{+5.6}_{-3.4}$ \\ \hline
 Control Galaxies & 198 & 105 & 93 & 297 & 138 & 159 & 363 & 177 & 186\\
 AGN & 34  & 13  & 21  & 40 & 11 & 29 & 55 & 23 & 32\\
AGN Fraction (\%) & 17.2$^{+3.0}_{-2.3}$ &  12.4$^{+3.9}_{-2.5}$ &  22.6$^{+4.9}_{-3.7}$ &  13.5$^{+2.2}_{-1.7}$ &  8.0$^{+2.9}_{-1.7}$ &  18.2$^{+3.4}_{-2.6}$ &  15.1$^{+2.1}_{-1.7}$ &  12.99$^{+2.9}_{-2.1}$ &  17.2$^{+3.1}_{-2.4}$ \\ \hline
AGN Enhancement & 1.06$^{+0.38}_{-0.26}$ &  1.15$^{+0.73}_{-0.40}$ &  1.00$^{+0.45}_{-0.30}$ &  0.98$^{+0.23}_{-0.34}$ &  1.36$^{+0.93}_{-0.48}$ &  0.83$^{+0.23}_{-0.37}$ &  1.20$^{+0.24}_{-0.31}$ &  1.70$^{+0.44}_{-0.61}$ & 0.84$^{+0.23}_{-0.36}$\\
\enddata
\tablecomments{z[a,b]: a\thinspace$<$\thinspace Redshift (z)\thinspace $<$\thinspace b.}
\label{table:enh_mor_mer_stern}
\end{deluxetable*}

\begin{figure*}[ht]
    \centering
     \hspace{-0.1in}
    \includegraphics[scale=0.67,trim = 1cm 1cm 1cm 1cm,clip=true]{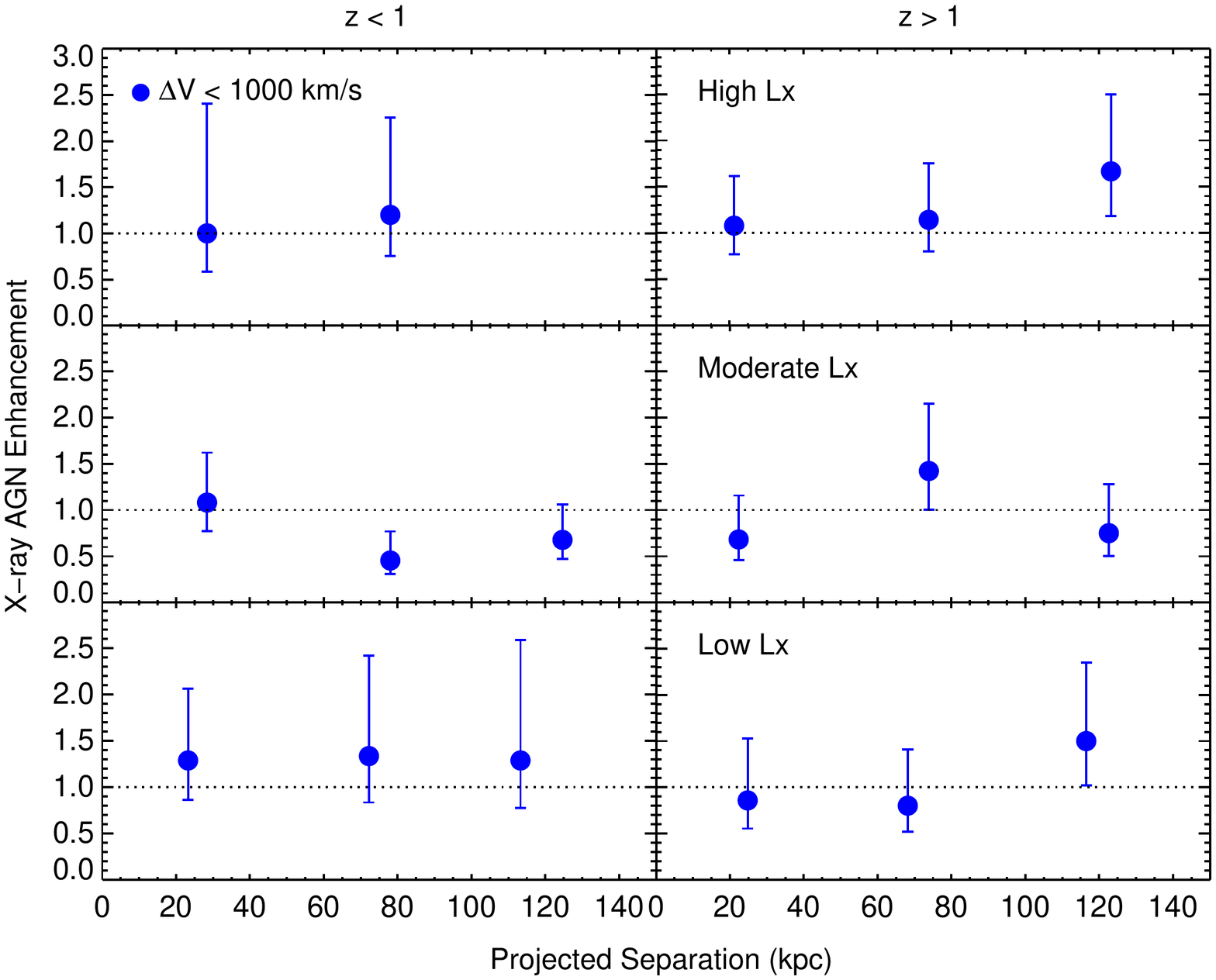}
    \caption{X-ray AGN enhancement as a function of projected separation for our sample of spectroscopically confirmed galaxy pairs with $\Delta V<1000$\thinspace km s$^{-1}$, divided into redshift and X-ray luminosity bins, as defined in Table~\ref{table:lz_cut} Figure \ref{fig:agn_lum_z}. The left and right panels correspond to $0.5<z<1.0$ and $1.0<z<3.0$, respectively. We see no significant AGN enhancement in any of our separation, redshift, or luminosity bins. At the highest separation in the high $L_{\rm{X}}$ bin at $z<1$ no point is plotted since there are no AGN in the paired galaxies satisfying these criteria. }

    \label{fig:enh_ladder_zsplit}
\end{figure*}

To investigate the level of interaction-induced X-ray AGN enhancement at different redshift epochs, we calculate the X-ray AGN enhancement in two redshift bins at the median redshift ($z\sim$1) of our spectroscopic pair sample: low $z$ ($z<1$) and high $z$ ($z>1$) bins. We show our results in Figure \ref{fig:enh_ladder_zsplit}, and find no statistically significant difference between the low $z$ and high $z$ AGN enhancement levels.

\begin{figure*}[ht]
    \centering
    \hspace*{-0.3in}
    \includegraphics[trim = 2cm 6cm 1.3cm 8.4cm,clip=true, scale=0.75]{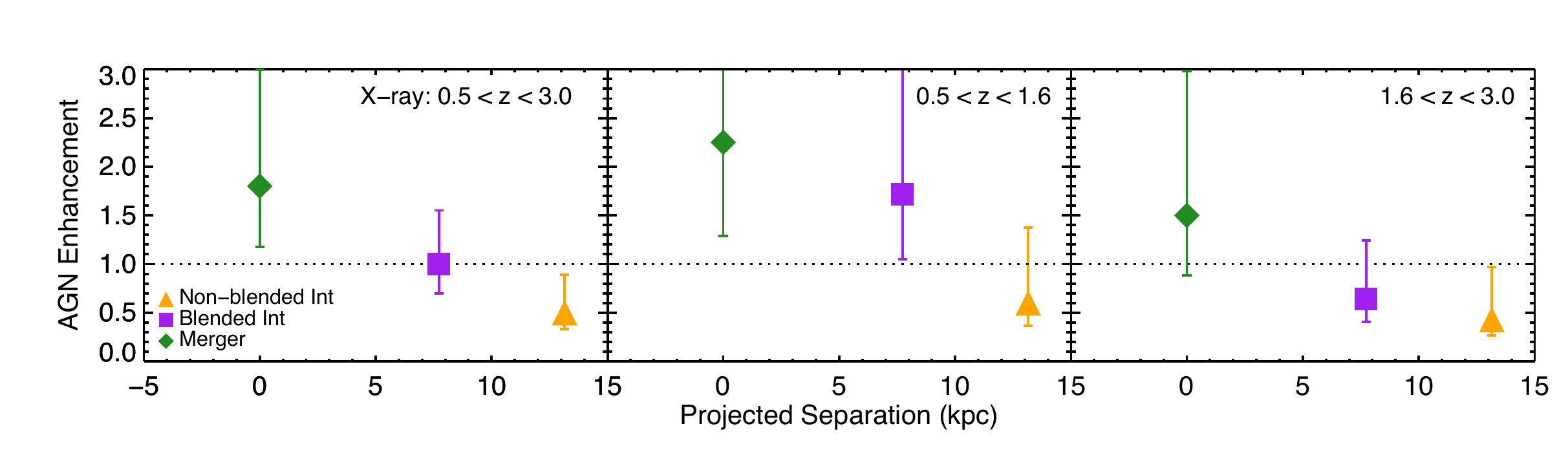}
    \caption{The level of X-ray AGN enhancement as a function of the  median projected separation for our visually identified mergers (filled green diamonds), blended interactions (filled purple squares), and non-blended interactions (filled orange triangles). The left, middle, and right panels correspond to the complete ($0.5<z<3.0$), low $z$ ($0.5<z<1.6$), and high $z$ ($1.6<z<3.0$) samples, respectively, with their values given in Tables~\ref{Fig:enh_mor_mer_xray}. The error bars on each point reflect the 1$\sigma$ binomial confidence limits, following the method of \citet{cameron2011}. The median redshift of all three visually identified samples combined is $\sim1.6$.}
    \label{Fig:enh_mor_mer_xray}
\end{figure*}

\subsection{AGN Enhancement in Visually Identified Interaction and Merger Sample}\label{sec:agn-enh-vis}
    We also analyze the level of AGN enhancement in our visually identified merger and interaction samples. We split the samples into two different redshift bins separated at the median redshift of the combined samples ($z\sim1.6$). We show our results for the X-ray AGN enhancement of the complete ($0.5<z<3.0$) merger and interaction samples as well as for the low $z$  and high $z$ samples in Figure \ref{Fig:enh_mor_mer_xray} and Table~\ref{table:enh_mor_mer_xray}. Though the number of AGN in the different redshift bins is small, and therefore the errors on the AGN enhancement value are large, we see a slight trend of increasing AGN enhancement with decreasing separation at all redshifts.  Additionally, the merger and blended interaction samples have smaller enhancement values at high $z$ compared to low $z$; however, the error bars are too large to make a statistically robust claim of redshift evolution.

\begin{figure*}[ht]
    \centering
    \hspace*{-0.3in}
    \includegraphics[trim = 2cm 6cm 1.3cm 8.4cm,clip=true, scale=0.75]{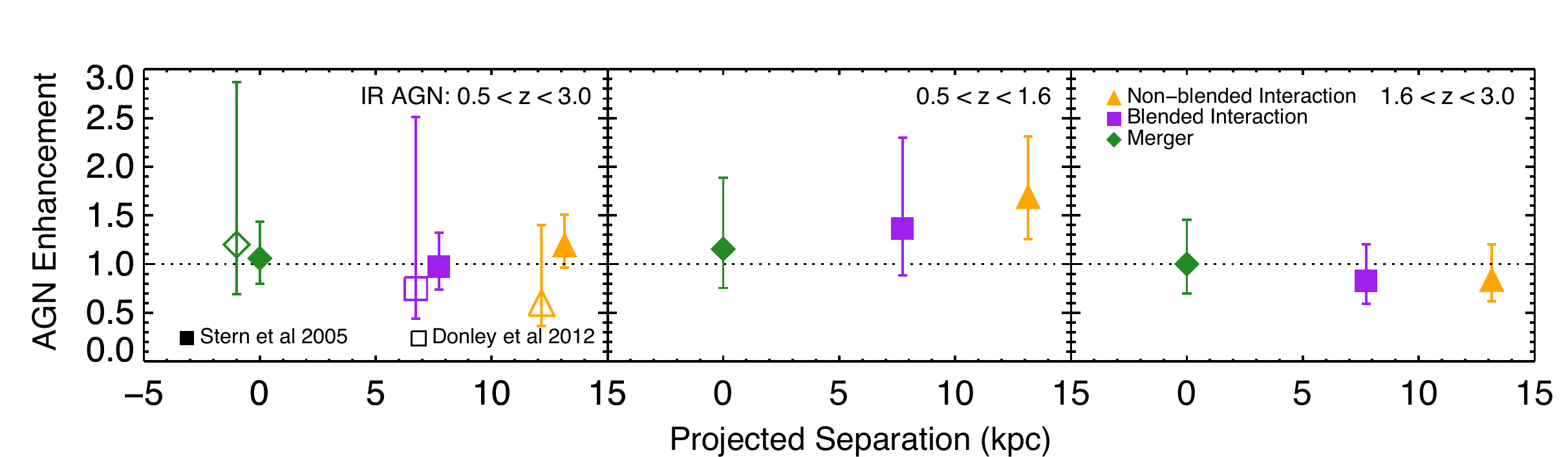}
    \caption{The level of IR AGN enhancement as a function of the median projected separation for our sample of visually identified mergers (green diamonds), blended interactions (purple squares), and non-blended interactions (orange triangles). The filled and open symbols correspond to IR AGN identified based on \citet{stern2005} and \citet{donley2012} criteria, respectively.  The left, middle, and right panels correspond to the complete ($0.5<z<3.0$), low $z$ ($0.5<z<1.6$), and high $z$ ($1.6<z<3.0$) samples, respectively, with their values given in Table~\ref{Fig:enh_mor_mer_ir}. The error bars on each point reflect the 1$\sigma$ binomial confidence limits, following the method of \citet{cameron2011}. The median redshift of the combined samples is $\sim1.6$.}

    \label{Fig:enh_mor_mer_ir}
\end{figure*}

We also calculate the IR AGN enhancement for the visually identified merger and interaction samples and show the results in Figure \ref{Fig:enh_mor_mer_ir} in the same redshift bins mentioned above. The \citet{donley2012} IR AGN enhancement values are presented in Table \ref{table:enh_mor_mer_ir}. As the number of AGN identified using these criteria is low, the error bars on the AGN enhancement value are large, and we do not see any enhancement. Since there is a larger number of AGN identified using the \citet{stern2005} criteria, the error bars are smaller. However, we do not see any enhancement for the full sample at any separation.  We further divide the Stern IR AGN enhancement values for the two redshift bins and find no significant level of enhancement overall at either redshift. In the low redshift bin, we see a slight enhancement for the non-blended interaction sample, which could indicate that enhancement is seen at an earlier stage of the merger process.

\section{Discussion} \label{sec:discussion}

To investigate the role of galaxy interactions and mergers on enhancing AGN activity at high redshift, we have compiled the largest known sample of major spectroscopically confirmed galaxy pairs at $0.5<z<3.0$, identified X-ray and IR AGN among them, and calculated the AGN fraction and level of AGN enhancement relative to a control sample of mass-, redshift-, and environment-matched isolated galaxies. We find that over this redshift range, major spectroscopic galaxy pairs, as well as visually identified interactions and mergers, do not show any statistically significant IR or X-ray AGN enhancement on average, except for the visually identified sample at the closest separations and those that have already coalesced into a single system. These results do not change significantly when the sample is split by X-ray luminosity.

\begin{figure*}[t]
    \centering
    \includegraphics[scale=0.81,angle=0, trim = 2.9cm 2.4cm 3cm 11cm,clip=true]{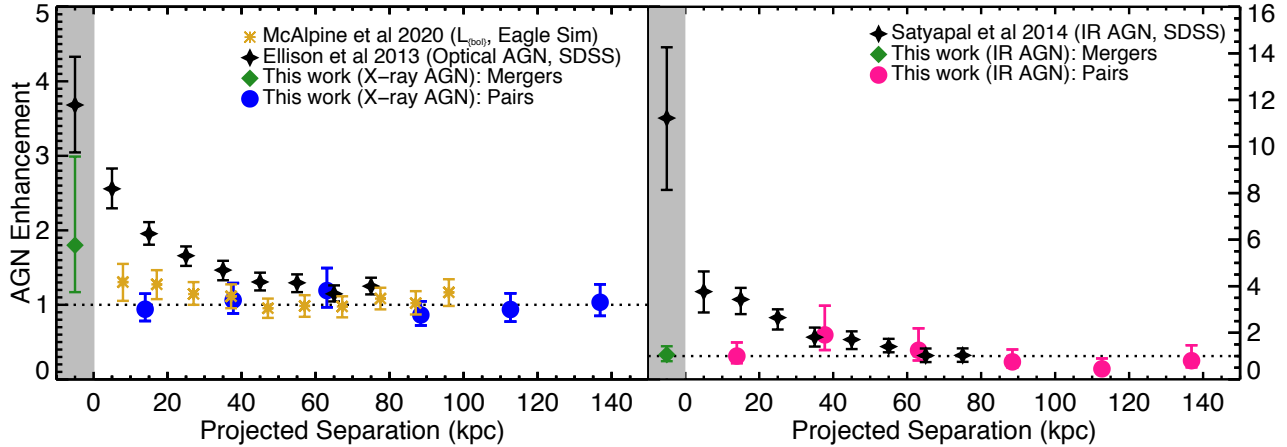}
   \caption{Comparison of our results with studies of galaxy pair samples in the local universe. Left: X-ray AGN enhancement as a function of projected separation for our sample of paired galaxies with $\Delta V<1000$ km s$^{-1}$ at $0.5<z<3.0$  (filled dark blue circles) and the visually identified merger sample (filled green diamond) in comparison with the results of \citet{ellison2013b} for optical AGN in SDSS spectroscopic paired galaxies and post mergers (filled black stars) at $0.01<z<0.20$ and the results of \citet{mcalpine2020} AGN ($L_{\rm{bol}}>2 \times 10^{42}$ erg s$^{-1}$) in pairs at $0.05<z<0.10$ from the cosmological simulation \textsc{EAGLE} (golden asterisks).  Right: IR AGN enhancement as a function of projected separation for our sample of paired galaxies with $\Delta V<1000$ km s$^{-1}$ (filled deep pink circles) and the visually identified merger sample (filled green diamond), based on the \citet{donley2012} criteria, in comparison with the results of \citet{satyapal2014} for IR AGN selected from WISE in SDSS spectroscopic paired galaxies and post mergers (filled black stars). The gray shaded region in both panels corresponds to merging/post-merger systems. All spectroscopic pairs correspond to major interactions (mass ratio $< 4$).}
    \label{fig:comp_comb}
\end{figure*}

Most studies in the nearby universe ($z\sim0$) find significant AGN enhancement in merging and/or interacting galaxies \citep[e.g.,][]{alonso2007, woods2007, ellison2011,carpineti2012, ellison2013b, satyapal2014, weston2017, fu2018, ellison2019}. For low redshift major galaxy pairs (stellar mass ratio $<4$) at $0.01<z<0.20$ selected from the SDSS, \citet{ellison2013b} find a clear trend of increasing optical-AGN excess (or enhancement) with decreasing projected separation ($<40$\thinspace kpc) as shown in the left panel of Figure \ref{fig:comp_comb}. They computed the largest enhancement of a factor of $\sim2.5$ at the closest projected separation ($<10$\thinspace kpc). Their estimate of the AGN enhancement for pairs with projected separation between 10 kpc and 20 kpc is 1.95$^{+0.16}_{-0.15}$, which is $\sim4.9\thinspace\sigma$ higher than our enhancement value for pairs ($V<1000$\thinspace km s$^{-1}$) with projected separation between 0 and 25 kpc (median $\sim$ 14 kpc) at $0.5<z<3.0$. While their post merger enhancement is higher than our value, it is almost within error bars. While the overall size of the interaction and merger samples likely plays a part in the difference between the enhancement across redshifts, the differences in how the samples were selected may also impact the results.

 For the same SDSS pairs and post merger sample as \citet{ellison2013b}, \citet{satyapal2014} use IR observations from the Wide-field Infrared Survey Explorer (WISE) all-sky survey to estimate IR AGN enhancement as shown in the right panel of Figure \ref{fig:comp_comb}. They identify IR AGN using the WISE color selection criteria of \citet{stern2012}. They also find increasing IR AGN enhancement with decreasing separation at $<$ 40 kpc, with the highest enhancement value of $\sim$ $4-6$ for pairs with projected separation of less than 10 kpc. Their IR AGN enhancement for pairs with projected separation between 10 kpc and 20 kpc is 3.43$^{+0.64}_{-0.63}$. It is $\sim3.8\thinspace\sigma$ higher than our IR AGN enhancement value of 1.00$^{+0.58}_{-0.31}$ for pairs with projected separation between 0 and 25 kpc (median $\sim$ 14 kpc). They also estimate an enhancement of 11.2$^{+3.1}_{-3.0}$ for their post-merger sample, which is $\sim3.3\thinspace\sigma$ higher than the IR AGN enhancement of 1.2$^{+1.6}_{-0.5}$ for our merger sample. Their result is $\sim2.5\thinspace\sigma$ higher than the optical AGN enhancement result for the same merger sample \citep{ellison2013b}.

The SDSS galaxy pair sample has a stricter relative velocity cut ($\Delta V<300$\thinspace km s$^{-1}$)  compared to our work (5000\thinspace km s$^{-1}$, 1000\thinspace km s$^{-1}$, and 500 km s$^{-1}$). However, our results do not show a significant enhancement for the $\Delta V<500$\thinspace km s$^{-1}$ pair sample at projected separation less than 25 kpc as shown in the left panel of Figure \ref{fig:enh_comb}.

\begin{deluxetable*}{lccc}
\tablewidth{0pt}
\tablecaption{IR AGN Enhancement (\citealt{donley2012} Criteria) for Visually Identified Mergers and Interactions: Left panel of Figure~\ref{Fig:enh_mor_mer_ir}}

\tablehead{\colhead{} & \colhead{Merger} & \colhead{Blended Interaction} & \colhead{Non-blended Interaction}}
\startdata
 Galaxies & 66 & 99 & 121 \\
 AGN & 2 & 1 & 2 \\
AGN Fraction (\%) & 3.0$^{+3.7}_{-1.0}$  & 1.0$^{+2.2}_{-0.3}$ & 1.7$^{+2.1}_{-0.5}$ \\ \hline
 Control Galaxies & 198 & 297 & 363 \\
 AGN & 5 & 4 & 10 \\
AGN Fraction (\%) & 2.5$^{+1.6}_{-0.7}$ & 1.3$^{+1.0}_{-0.4}$ & 2.8$^{+1.1}_{-0.6}$ \\ \hline
AGN Enhancement & 1.20$^{+1.67}_{-0.51}$ & 0.75$^{+1.76}_{-0.31}$ & 0.60$^{+0.80}_{-0.23}$\\
\enddata
\tablecomments{Merger, Blended Interaction and Non-blended Interaction are defined based on \citet{kartaltepemor15} (see Section \ref{sec:visual_int_mer}).}
\label{table:enh_mor_mer_ir}

\end{deluxetable*}

While in the nearby universe $\sim$80\% of all quasars (or high luminosity AGN) show signs of a recent or ongoing merger \citep{sanders1988a,sanders1988b,bannert2008,urrutia2008}, our results do not show AGN enhancement even in the highest X-ray luminosity range.  Our results are consistent with the results of \citet{marian19}, who consider the highest specific accretion broad line AGN at the peak epoch of AGN activity around $z\sim2$ and find no significant difference in the merger fraction of the AGN-host galaxies and (mass- and redshift-matched) non-AGN galaxies. However, \citet{treister2012} find that mergers are responsible for triggering the highest luminosity AGN at $0<z<3$ ($z<1$ for most of their sample), with no signs of redshift dependence. One possible explanation for this difference is that our work on spectroscopic pairs probes the earliest stages of the merger process, while galaxies are still distant pairs, rather than the most advanced stage mergers expected to fuel quasars, and our visually identified merger and interaction samples are too small to make a statistically significant claim.

One of the main differences between many local studies and our study is the method used to identify AGN. Most of these local studies use optical AGN selected using emission line ratios while we use X-ray and IR observations to identify AGN. Since it is possible that AGN would be visible at different wavelengths at different stages of the merger sequence, due to factors such as dust obscuration, there could be inherent differences between the level of AGN enhancement calculated based on different AGN identification methods. Furthermore, the relative timescale of AGN triggering and the merging process, as well as the duration of AGN activity, could also change with redshift, resulting in differences in AGN enhancement at high and low redshifts \citep{mcalpine2020}. However, we note that comparison between our IRAC-selected IR AGN with WISE-selected IR AGN among local pairs \citep{satyapal2014}, shown in Figure 15, highlight the difference between local and high redshift interacting systems for similar types of AGN.

 \citet{silverman2011} present a sample of 562 galaxies in kinematic pairs $(0.25<z<1.05$, $1<$\thinspace mass ratio \thinspace $<10)$ and find a higher (by a factor of 1.9) AGN fraction in paired galaxies at projected separations less than 75\thinspace kpc (relative line-of-sight velocity less than 500 \thinspace km s$^{-1}$) compared to their control sample of galaxies. We note that since their sample was based on zCOSMOS observations, their major (mass ratio $<4$) pairs are included as a subset of the ones used for our study. However, our results are not in strong agreement.

 The control sample used by \citet{silverman2011} consists of the non-paired galaxies in their survey and the same sample is used for different separation bins. Based on K-S tests, they claim that there is no difference between the mass distribution of pairs and controls in projected separation, line-of-sight velocity, and redshift bins. Environmental effects on larger scales can also play a role in AGN fueling. Using a mock catalog of an SDSS-like survey, \citet{perez2009} show that although mass is likely the most crucial parameter to match while generating a control sample to study the effect of galaxy interactions, by matching in both redshift and environment the differences between the pairs and control sample are reduced by 70\%. \citet{ellison2013b} find that the main reason they were able to estimate AGN excess at larger separations compared to \citet{ellison2011} is the addition of environment-matching of controls. Hence, it is likely critical to control for environment as well. Our controls were carefully matched to each paired galaxy to account for any subtle variations in mass, redshift, and environment of the general galaxy population, enabled by the ever-growing set of spectroscopic observations in these fields.

 \citet{silverman2011} also include both major and minor interactions while our work focuses on just major interactions. This should affect the results, though one would expect that this would have an effect in the opposite direction to what we see (major interactions should see a stronger enhancement that minor interactions). Studies in the local universe show that the effect of minor interactions on AGN activity could be different from that of major interactions \citep{ellison2011}. Further work at high redshift is required to determine the impact of minor mergers. We plan to explore these differences in a future paper.

 We also compare our AGN enhancement results to those using the sample of kinematic pairs selected by \citet{mantha2018} in the CANDELS fields. Applying the same cuts to their sample as we used for our pairs results in a total sample size of 154 pairs with $\Delta V<500$\thinspace km\thinspace $^{-1}$ and projected separations of $5-150$\thinspace kpc. Unfortunately, there are too few pairs in the closest separation bin and too few control galaxy candidates to conduct a fair comparison with our sample. Note that this pair sample is almost an order of magnitude smaller than ours (we have 1066 pairs with $\Delta V<500$\thinspace km\thinspace $^{-1}$) because we included the larger 2\thinspace deg$^2$ COSMOS field, our own DEIMOS, GMOS, and MOSFIRE observations, and spectroscopic samples in these fields have generally grown since their study was first published. Since these were selected within CANDELS, the \citet{mantha2018} pair sample is a subset of the pairs included in our analysis. This highlights the importance of using large spectroscopic samples for this analysis.

 As discussed above, generating a well-matched control sample is one of the crucial parts of this analysis. Here, we highlight different factors that play a significant role in how controls are selected. One of the main limiting factors is the availability of spectroscopic redshifts. Redshift completeness falls off as a function of redshift due to the availability of spectral lines in observable wavelength ranges and the increasing faintness of galaxies at high redshift. This biases the sample toward pairs at lower redshift and the spectroscopic incompleteness results in missing pairs. An effect of this is that the control sample could contain galaxies that are actually in a pair, but we are missing the redshift for its companion. This could result in a dilution of the measured AGN enhancement, particularly at high redshift. Similarly, some galaxies in the control sample may be at an advanced merging stage and missed by our selection. We attempted to account for this by removing the visually identified mergers and interactions from the control parent sample, but since that selection was fairly conservative, there are almost certainly many mergers that have been missed and could have been included in the control sample.

 It is also important to note that any biases and selection effects present in the spectroscopic redshift samples will be present in our pair sample. Spectroscopic surveys in these fields are inhomogenous overall and each survey has a different goal in mind for targeting. Of particular note, the spectroscopic completeness of X-ray AGN is higher than the general galaxy population in these fields since there have been many campaigns to specifically target X-ray AGN. We attempt to mitigate this by requiring all controls to have spectroscopic redshifts and all controls to come from the same field as the galaxy pairs so that any selection effects are present in both samples. Therefore, we expect that these selection effects have minimal impact on our final AGN enhancement results.

 While our kinematic pair sample is not affected by the dimming of low surface brightness features at high redshift, our sample of visually identified interactions and mergers certainly are. The observational bias of surface brightness dimming results in a decrement of three magnitudes in sensitivity from $z=0$ to $z=1$. Despite using deep \textit{HST} images to visually identify the interaction and merger samples, these samples are incomplete as many interaction features at high redshift are too faint to be identified. In addition to being difficult to identify, many classifiers may disagree on the presence of merger signatures, due to their faintness as well as to the fact that other physical processes can be responsible for morphological disturbances at high redshift. Our selection in this paper is intentionally conservative -- all of the galaxies identified as mergers and interactions have a high level of confidence due to the presence of strong signs of disturbance. Therefore, this analysis is certainly insensitive to all of the mergers in these fields and our resulting sample is very small, affecting our statistics. This could result in some missing mergers being included in our control sample, diluting any AGN enhancement in our measurement.

 We compare our results for our visually identified samples with the results of \citet{lackner2014}. They apply an automated method of identifying mergers by median-filtering the high-resolution COSMOS \textit{HST} images to distinguish two concentrated galaxy nuclei at small separations, i.e., to identify late-stage mergers at $0.25<z<1.0$, and also used X-ray observations to identify AGN. They find that their late-stage merger sample has higher X-ray AGN activity by a factor of $\sim2$ compared to their mass- and redshift-matched control sample. Our results for the visually classified merger sample are consistent within the error bars of these results.

To study the effect of using different criteria to define merger and interaction samples, we also calculate the level of AGN enhancement for a redefined sample of interacting and merging galaxies based on the criteria of \citet{rosario2015} applied to the full visual classification catalog of \citet{kartaltepemor15}. \citet{rosario2015} assign an interaction metric (IM) value for each visual classification of an object. The IM value ranges from IM = 0 (a clearly undisturbed object with no obvious nearby companion) to IM = 1 (an obvious late-stage merger). The intermediate IM values of IM = 0.25 is assigned to objects in apparent pair or multiple systems (with a maximum separation of several arcseconds apart) with no clear signs of interaction, which may or may not be associated to each other, IM = 0.5 for non-blended interactions, i.e., systems with apparent interaction signs with galaxies in different {\it H}-band segmentation maps, and finally IM = 0.75 is assigned to blended interactions, i.e., distinct interacting galaxies that share a segmentation map. Based on the average IM (averaged over all the classification IMs), \citet{rosario2015} define interaction classes as: $0.0\le IM \le 0.2$ for Isolated, $0.2 < IM \le 0.5$ for interacting, and $0.5 < IM \le 1.0$ for mergers. Therefore, everything with a visual classification is divided into these three classes. These classes are more liberally defined than our constraints. For example, if we have a galaxy for which each classifier agrees about its classification as a `blended interaction,'  it would be included in the `Merger' (not interaction) class of the \citet{rosario2015} classification metric.

Applying this metric to the \citet{kartaltepemor15} catalog in all five CANDELS fields, and applying our mass and redshift cuts, we identified 518 mergers, 2120 interactions, and 4606 isolated galaxies. We match control galaxies for these objects using photometric redshifts (following the same method that is used for our visually identified interaction and merger samples). We calculate an X-ray AGN enhancement of 1.07$^{+0.22}_{-0.17}$ and 0.80$^{+0.08}_{-0.07}$ for their merger and interaction samples, respectively. While the error bars are smaller due to the larger sample identified this way, the result agrees overall with our sample discussed above. Hence, we do not find significant AGN enhancement in this more inclusive merger and interaction sample.

Another approach to understanding the effect of galaxy interactions on AGN activity is to use simulations of galaxy mergers. Most simulations of galaxy mergers between nearby massive gas-rich galaxies show enhancement in both AGN activity and star formation rate caused by interaction induced gravitational torques \citep[e.g.,][]{barnes1991,mihos1996,hopkins2009}. However, for high redshift galaxy interactions and mergers, simulations find a varying range of results.

\citet{mcalpine2020} conducted a study over a large redshift range that uses a similar approach to ours using the cosmological hydrodynamical \textsc{eagle} simulation. They find a higher AGN fraction in galaxies with close major companions relative to their controls. As shown in the left panel of Figure \ref{fig:comp_comb}, for AGN identified based on a bolometric luminosity cut ($L_{rm{bol}}>10^{42}$ erg s$^{-1}$), they see an enhancement of 1.28$^{+0.23}_{-0.21}$ at projected separation of $\sim$ 15 kpc at $0.05<z<0.10$, which is within the error bars of our X-ray AGN enhancement value (0.94$^{+0.21}_{-0.16}$) for projected separation $<25$ \thinspace kpc at $0.5<z<3.0$. However, for AGN defined based on an Eddington rate cut, they see a strong trend of increasing AGN excess with decreasing projected separation starting at 3D separations of $50-100$ \thinspace kpc for $z<2$ galaxies with the highest excess value of $1.2-1.3$ at 10 \thinspace kpc. They defined redshift bins of $0<z<1$, $1<z<2$, $2<z<5$, and find a decreasing AGN enhancement with increasing redshift. For $z>1$ for both AGN definitions, they find excess values oscillating around $1.2-1.3$.

\citet{mcalpine2020} also show the effect of different ways of selecting controls matched to a range of different parameters and their combinations: mass, redshift, environment, gas mass, BH mass, and halo mass. They find that the AGN excess value decreases when the number of matched parameters increases with a deviation within a factor of two. Furthermore, they find that results based on the Eddington luminosity criteria were more sensitive to the control matching compared to the results based on the bolometric luminosity criteria. They also find that the trend of increment in AGN excess with decreasing separation is not affected by the change in the matching criteria.

We control for mass, redshift, and environment, and our results do not show any significant AGN enhancement for the paired galaxies. For the visually identified sample there are hints of slight X-ray (Figure \ref{Fig:enh_mor_mer_xray}) and IR AGN (Figure \ref{Fig:enh_mor_mer_ir}) enhancement at $0.5<z<1.6$ with very low ($<1.5\thinspace\sigma$) statistical significance. These results suggest that there might be redshift evolution in the effect of interactions and mergers on AGN activity, even at $z<1$. As suggested by simulations, the interaction and merger induced gas inflows responsible for the enhancement in AGN activity could strongly depend on the properties of the galaxies, such as their gas fractions \citep{cox2008,dimatteo2008,fensch2017}.  The gas fraction in massive spiral galaxies increases from  $\sim$10\% at low redshift ($z\sim0$) to $\sim$ 50\% at high redshift  \citep[z $\sim$ 2,][]{daddi2010,tacconi2010,scoville2014}. Furthermore, gravitational instabilities, and hence velocity dispersion, are also higher ($\sigma\sim40$\thinspace km s$^{-1}$) at high redshift compared to low redshift ($\sigma\sim10$\thinspace km s$^{-1}$) \citep{stott2016}. This may weaken the strong inflows, essential for the enhancement in AGN activity. The efficiency of galaxy interactions and mergers in enhancing the AGN activity could thus be weaker at high redshift compared to low redshift.

\section{Summary} \label{sec:summary}

In this paper, we investigate the effect of galaxy interactions on AGN activity using deep multiwavelength observations from the CANDELS and COSMOS surveys. We generated the largest known sample of 2381 major spectroscopic galaxy pairs with $\Delta V<5000$ km s$^{-1}$ over $0.5<z<3.0$, with the stellar mass of both galaxies greater than 10$^{10}\thinspace$M$_{\odot}$ and with the stellar mass ratio of the primary (more massive) to the secondary (less massive) galaxy less than four. We also selected samples of visually identified interactions and mergers consisting of 61 galaxy pairs of non-blended interactions, 100 galaxy pairs of blended interactions, and 66 galaxy mergers.

To compute the interaction-induced AGN enhancement, we generate a stellar mass-, redshift-, and environment-matched control sample of three galaxies for each paired galaxy and visually identified interaction and merger selected from the same field. We define the AGN enhancement as the ratio of the AGN fraction of the paired or visually identified galaxy samples to that of the corresponding control galaxy sample.

We explored the effect of using different relative line-of-sight velocity cuts by constructing samples with three different cuts: $\Delta V<500$ km s$^{-1}$ (1066 pairs), $\Delta V<1000$ km s$^{-1}$ (1345 pairs), and $\Delta V<5000$ km s$^{-1}$ (2381 pairs). We do not see significant AGN enhancement for any of these samples; the results of all three are consistent within error bars.

For the closest projected separation bin ($<25$\thinspace kpc, median $\sim$ 14\thinspace kpc) in our sample ($0.5<z<3.0$, $\Delta V<1000$\thinspace km s$^{-1}$), we find enhancements of a factor of 0.94$^{+0.21}_{-0.16}$ and 1.00$^{+0.58}_{-0.31}$ for X-ray and IR-selected AGN, respectively. These results appear to be somewhat in contrast with $z \sim 0$ results that indicate strong AGN enhancement in the closest pairs, as shown in Figure \ref{fig:comp_comb}. At roughly equivalent small separations ($\sim$ 15 kpc), our X-ray enhancement result is $\sim4.9\thinspace\sigma$ lower than the local optical AGN enhancement \citet{ellison2013b}, and our IR AGN enhancement is $\sim3.8\thinspace\sigma$ lower than local IR AGN enhancement \citep{satyapal2014}. While the X-ray and optical AGN enhancement results for merger samples are almost within error bars, our IR AGN enhancement is $\sim$ 3.3 sigma lower than the local result. These discrepancies suggest that high redshift mergers and interactions might be less efficient at triggering AGN compared to such interactions at low redshift as also suggested by some simulations \citep[e.g.,][]{fensch2017,mcalpine2020}.

Considering the different depth of X-ray observations in the CANDELS and COSMOS fields, we also apply different redshift and luminosity cuts to account for X-ray completeness and to conduct a consistent analysis among all the fields.  We further divide our sample at its median redshift of $\sim$ 1.0 to compare the enhancement results in the low redshift ($0.5<z<1.0$) and high redshift ($1.0<z<3.0$) halves of the sample. We find no significant enhancement in AGN activity in any of our pair separation, redshift, or X-ray luminosity bins in our galaxy pairs and visually identified mergers relative to the control sample of galaxies.

The error bars on our results are large enough to hide possible low-level AGN enhancement. A larger sample of pairs across a wide range in redshift is needed, especially at smaller separations, to make statistically significant claims about AGN enhancement level differences at high and low redshifts. In the upcoming decade, surveys using facilities such as the {\it James Webb Space Telescope} (JWST), the Vera Rubin Observatory, {\it Euclid}, and the {\it Nancy Grace Roman Space Telescope},  along with follow-up spectroscopic and multiwavelength broad band observations (e.g., X-ray observations from eROSITA) will help to improve the statistics and enable a quantitative determination of how galaxy interactions and mergers affect AGN activity over cosmic time.

\section*{Acknowledgements}

We thank the anonymous reviewer for their critical reading of the manuscript and providing detailed and insightful comments, which greatly enhanced the quality and clarity of the paper. Support for this work was provided by NASA through grants HST-GO-13657.010-A and HST-AR-14298.004-A awarded by the Space Telescope Science Institute, which is operated by the Association of Universities for Research in Astronomy, Inc., under NASA contract NAS 5-26555. Support was also provided by NASA through grant NNX16AB36G as part of the Astrophysics Data Analysis Program. This work was also supported by start-up funds and the Dean's Research Initiation Grant fund from the Rochester Institute of Technology's College of Science. ES would like to thank Athis Osathapan, Krystal Tyler, and Kevin Cooke for intriguing discussions and assistance with spectral energy distribution fitting, which was performed using the computational resources and support from Research Computing Services at the Rochester Institute of Technology \citep{https://doi.org/10.34788/0s3g-qd15}. ES thanks the LSSTC Data Science Fellowship Program, which is funded by LSSTC, NSF Cybertraining Grant \#1829740, the Brinson Foundation, and the Moore Foundation; her participation in the program has benefited this work. HI acknowledges support from JSPS KAKENHI Grant Number JP19K23462. ET acknowledges support from FONDECYT Regular 1190818, ANID PIA ACT172033 and Basal-CATA AFB170002.

Some of the data presented herein were obtained at the W. M. Keck Observatory, which is operated as a scientific partnership among the California Institute of Technology, the University of California and the National Aeronautics and Space Administration. The Observatory was made possible by the generous financial support of the W. M. Keck Foundation. Based in part on observations obtained at the international Gemini Observatory and processed using the Gemini IRAF package \citep{tody1986,tody1993}, a program of NOIRLab, which is managed by the Association of Universities for Research in Astronomy (AURA) under a cooperative agreement with the National Science Foundation. on behalf of the Gemini Observatory partnership: the National Science Foundation (United States), National Research Council (Canada), Agencia Nacional de Investigaci\'{o}n y Desarrollo (Chile), Ministerio de Ciencia, Tecnolog\'{i}a e Innovaci\'{o}n (Argentina), Minist\'{e}rio da Ci\^{e}ncia, Tecnologia, Inova\c{c}\~{o}es e Comunica\c{c}\~{o}es (Brazil), and Korea Astronomy and Space Science Institute (Republic of Korea). The authors wish to recognize and acknowledge the very significant cultural role and reverence that the summit of Maunakea has always had within the indigenous Hawaiian community.  We are most fortunate to have the opportunity to conduct observations from this mountain.

This work is also based in part on observations made with the NASA/ESA Hubble Space Telescope, obtained from the Data Archive at the Space Telescope Science Institute, which is operated by the Association of Universities for Research in Astronomy, Inc., under NASA contract NAS 5-26555, observations made with the {\it Spitzer Space Telescope}, which is operated by the Jet Propulsion Laboratory, California Institute of Technology under a contract with NASA, and observations made by the {\it Chandra X-ray Observatory} and published previously in cited articles.

\software{LePhare \citep{arnouts2002,ilbert2006}, MAGPHYS \citep{dacunha2008}, SpecPro \citep{masters2011}, spec2d IDL pipeline \citep{newman2013,cooper2012a}, IRAF \citep{tody1986,tody1993}}
\bibliography{main.bib}
\end{document}